\documentclass[draft]{agujournal2019}
\usepackage{url} 
\usepackage{lineno}
\usepackage[inline]{trackchanges} 
\usepackage{soul}
\usepackage{amssymb}
\usepackage{booktabs}
\usepackage{subcaption}
\usepackage{algorithm}
\usepackage{graphicx}
\usepackage{algpseudocode}
\usepackage{mathtools}
\usepackage{xcolor}

\linenumbers
\draftfalse

\journalname{AGU}

\begin{document}
\nolinenumbers
%
%


\title{Accurate Long-term Air Temperature Prediction with a Fusion of Artificial Intelligence and Data Reduction Techniques}

%
%




\authors{Du\v{s}an Fister\textsuperscript{1}, Jorge P\'erez-Aracil\textsuperscript{2}, C\'esar Pel\'aez-Rodr\'iguez\textsuperscript{1}, Javier Del Ser\textsuperscript{3, 4}, Sancho Salcedo-Sanz\textsuperscript{1}}

\affiliation{1}{Department of Signal Processing and Communications. University of Alcala, 28805, Madrid, Spain.}
\affiliation{2}{Department of Computer Systems Engineering, Universidad Polit\'ecnica de Madrid, 28038, Madrid, Spain.}
\affiliation{3}{TECNALIA, Basque Research \& Technology Alliance (BRTA), 48160 Derio, Spain}
\affiliation{4}{University of the Basque Country (UPV/EHU), 48013 Bilbao, Spain}





\correspondingauthor{J. P\'erez-Aracil}{jorge.perez.aracil@upm.es}




\begin{keypoints}
\item Artificial intelligence models may obtain accurate long-term air temperature prediction.
\item Convolutional neural networks can efficiently process geospatial data to obtain good temperature forecasts.
\item Several data reduction methods and feature selection can be employed to improve the results on air temperature forecasting.
\end{keypoints}

%
%

%
%


\begin{abstract}
In this paper three customised Artificial Intelligence (AI) frameworks, considering Deep Learning (convolutional neural networks), Machine Learning algorithms and data reduction techniques are proposed, for a problem of long-term summer air temperature prediction. Specifically, the prediction of average air temperature in the first and second August fortnights, using input data from previous months, at two different locations, Paris (France) and Córdoba (Spain), is considered. The target variable, mainly in the first August fortnight, can contain signals of extreme events such as heatwaves, like the mega-heatwave of 2003, which affected France and the Iberian Peninsula. Thus, an accurate prediction of long-term air temperature may be valuable also for different problems related to climate change, such as attribution of extreme events, and in other problems related to renewable energy. The analysis carried out this work is based on Reanalysis data, which are first processed by a correlation analysis among different prediction variables and the target (average air temperature in August first and second fortnights). An area with the largest correlation is located, and the variables within, after a feature selection process, are the input of different deep learning and ML algorithms. The experiments carried out show a very good prediction skill in the three proposed AI frameworks, both in Paris and Córdoba regions. 
\end{abstract}


%
%

%


%
%
%
%

\section{Introduction}
Seasonal Climate Prediction (SCP) has gained momentum in the last decade \cite{doblas2013seasonal}, becoming an important field of study, with applications in very different areas such as agriculture, risk management, long-term energy planning or climate change and extreme events modelling \cite{pepler2015ability,salcedo2022analysis}, among others. SCP problems are specially interesting in the current context of climate change, since they may have important consequences in the future \cite{masson2021climate}. One of such climate change effects are the constantly rising long-termed average temperature, coined as the so-called global warming, and associated greenhouse gases \cite{seager2019strengthening,change2018global}. However, the constantly changing weather conditions not only affect the long-term temperature averages, but also stipulate temporally much shorter periods with drastically large deviations from steady levels, producing extreme phenomena such as heatwaves and severe droughts.

Evidence shows that these extreme weather events can cause worldwide consequences and impacts in natural resources (agriculture, construction, energy) \cite{bergmann2016natural}, financial sector \cite{wolf2010social} and of course  human's health \cite{diaz2002effects,diaz2002heat}. Also, one of the effects of climate change is to produce warmer summers \cite{pena2015multidecadal}, which can be further studied by predicting average summer months temperature at a long-term basis. SCP related to air temperature are, therefore, extremely important and challenging problems, due to the long-term prediction time-horizon involved in these problems. There are many different previous works involving problems related to long-term prediction of air temperature, many of them involving Machine Learning (ML) or Artificial Intelligence (AI) methods. For example, there have been several previous works discussing the application of Neural Networks to long-term air temperature prediction problems, such as in \cite{ustaoglu2008forecast}, where three different types of neural networks were applied to a problem of daily mean, maximum and minimum temperature time
series in Turkey. In \cite{abdel1995modeling} different artificial neural networks were applied to a problem of daily maximum temperature prediction in Dhahran, Saudi Arabia. Data for 18 weather parameters were considered as input variables, and the objective was to predict the maximum temperature on a given day, with different prediction time-horizons up to 3 days in advance. In \cite{de2009artificial} a multi-layer perceptron neural network is applied to the prediction of the maximum air temperature in the summer monsoon season in India. The mean temperature oft previous months in the period of analysis is considered as inputs for the system.

Other ML approaches have also been applied to long-term prediction of air temperature. For example, in \cite{paniagua2011prediction} a Support Vector Regression algorithm (SVR) was applied to a problem of daily maximum air temperature prediction, with a 24h prediction time-horizon. Input variables such as previous air temperature, precipitation, relative humidity and air pressure and synoptic situation were considered. Results in different European measurement stations were reported. In \cite{mellit2013least} a least squares SVR algorithm is applied to prediction of time series temperature in Saudi Arabia. In \cite{ahmed2020multi} different ML approaches are proposed to develop multi-model ensembles from global climate models. The objective is to obtain annual prediction of monsoon maximum temperature and minimum temperature, among other variables, over Pakistan. In \cite{peng2020prediction} two ML algorithms (MLP and natural gradient boosting (NGBoost)), are applied to improve the prediction skills of the 2-m maximum air temperature, with prediction time horizon with lead times from 1 to 35 days. In \cite{Oettli2022} a number of ML algorithms such as neural networks, SVMs, RF, Gradient Boosting or regression trees have been applied to the prediction of surface air temperature two months in advance, with input data two months in advance from SINTEX-F2, a dynamical prediction system. Results in data from Tokio (Japan) have confirmed the good skill of the prediction. 

In the last years, Deep Learning (DL) algorithms have been successfully applied to long-term air temperature prediction problems, such as in \cite{karevan2020transductive}, where a type of LSTM network (Transductive LSTM) is applied to a problem of temperature prediction in Belgium and the Netherlands, or \cite{vos2021long} where a coupling of CNN and LSTM (ConvLSTM) is proposed for a long-range air temperature prediction problem.
Note that such DL models get complex very soon, and in many cases the training sample size needs to be extraordinary large. However, one important issue is that the training size is usually severely constrained, due to the availability of the historic data. There are several public meteorological databases from measurements or Reanalysis \cite{salcedo2020machine}, but many of them are limited to data from 1950 or 1979 such as Reanalysis data. This means that, in many cases there are 72 years of the meteorological data available effectively for the given geographical location and, if severe extreme events occur every 10-15 years, there are just a sample of extreme events incorporated within the data. The application of DL complex models to SCP problems implies, therefore, a trade-off with the data availability, in which improvements can be expected by means of information fusion \cite{rasp2018deep}. For example, in \cite{rasp2020weatherbench} (WeatherBench) an example of an image-to-image translation using the CNN (among other methods) for medium-range weather predictions of up to 5 days has been shown. In that paper, the inputs are organised as images, where each pixel represents a geographical location. Similarly, the outputs are as organised as images, hence the image-to-image translation. An improved WeatherBench approach with the pre-trained ResNet was proposed soon later in \cite{rasp2021data}. In \cite{jin2022deep} the application of CNN on the case study for climate prediction over China was shown. The so-called capsule neural networks (CapsNets) were proposed for DL analog predictions in \cite{chattopadhyay2020analog}, where they have exhibited significant statistical benefits compared to usual DL practices. In \cite{taylor2022deep} an integrated framework for predicting the sea surface temperature was proposed. The proposed method, so-called Unet-LSTM, was based on the LSTM showed mixed prediction skills for predicting two of the past extreme events, again on the image-to-image basis to emphasise the ``big-picture'' phenomena.

Based on the excellent performance previously shown by ML and DL approaches in air temperature prediction, in this paper we propose and analyze different ML and DL approaches with data fusion and data reduction techniques, for a  long-term air temperature prediction problem. Specifically, the objective of the research is to predict the average temperature of the first and second August fortnights, using meteorological data from previous months. This problem has different climatological and energy-related applications, such as detection and attribution of heatwaves or prediction of energy consumption, among others. In order to achieve this objective, we propose the following procedure, based on artificial intelligence techniques: we start with a first correlation analysis among predictive variables (meteorological variables) and the target variable (air temperature of the first and second August fortnights). This correlation analysis defines a Geographic Selection Area (GSA), a reduced area of study with the highest correlation among predictive and target variables. Following, we apply an Exhaustive Feature Search (EFS) to reduce the number of predictive (input) variables in the modelling methods. Then, three different computational frameworks for prediction are defined: First, we analyze the performance of a Convolutional Neural Network (CNN), with video-to-image translation. In this case, a video stands for a sequence of half-monthly climate data and a 3D CNN filters are exploited to reduce the input dimension to an output image. Here, pixels represent geographical coordinates and the 3-channelled RGB dimensions are replaced by $n$-channelled climate data. This CNN with video-to-image translation has been applied to the whole GSA defined data area. The second computational framework, independent of the first one, analyzes the performance of several ML approaches (Multi-linear regression (LR), Lasso regression, Regression trees (DT) and Random Forest (RF). In this case we have selected a single node of the GSA (the most correlated node), and the inputs to the ML approaches are time series of climate variables (not images). The last computational framework also considers the CNN as a central processing element, but instead of processing the raw data, it relies on a pre-processing step with Recurrence Plots (RPs) \cite{eckmann_recurrence_1987,thiel2004much}, which convert time series into images. In this case, RPs share the same initial data as the ML methods, i.e. a time series of length $t$ for a given geographic coordinate with the highest correlation with the target. Two different methodologies, i.e., analogue and binarised RPs are applied and compared. After the application of the RP, the resulting image is applied to a CNN in order to obtain a final air temperature prediction within this computational framework. 

The proposed methodology, with the three computational frameworks proposed are trained and validated on Reanalysis data (ERA5 Reanalysis), considering two different geographical locations in Europe: Paris (France, northern Europe) and Córdoba (Spain, Iberian peninsula), where episodes of extreme summer temperature have occurred in the last decades. For example, August's 2003 extreme temperature rise severely impacted south of Spain, such as the city of Córdoba. Also, another extreme temperature rise was recorded in August 2003 that severely shocked the north of the France \cite{garcia2010review}.  

The rest of the paper has been structured in the following way: next section discusses the different data handling and fusion techniques used in this paper. We discuss here the Reanalysis data used, the processing of the data into geographical area selection and a process of feature selection to obtain the best set of inputs data for the DL approaches. Section \ref{sec:methods} present the proposed CNN-based methods for accurate long-term air prediction. Section \ref{sec:Experiments} shows the performance of the proposed DL approaches based on CNN, in the two geographical areas considered (Paris and Córdoba). A comparison with alternative ML algorithms and a discussion of the findings are also shown in this section. Section \ref{sec:Conclusions} closes the paper with some final remarks and conclusions on the research work carried out.

\section{Data handling and data reduction techniques} \label{data}

Original meteorological data were obtained from a single source, the ERA5 Reanalysis \cite{hersbach2020era5}, compiled and maintained by European Centre for Medium-Range Weather Forecasts (ECMWF) \cite{ECMWF} in a GRIB file format. The considered input and output variables are listed in Table \ref{tab:vars}, and the corresponding notation are given as used throughout the paper. Each data variable were initially obtained on hourly basis, ranging from 1st January 1950 to 31st December 2021, between latitude and longitude coordinates ranging $[70^\circ \textmd{N},~20^\circ \textmd{N}]$ and $[30^\circ \textmd{W},~30^\circ \textmd{E}]$, with the coordinate resolution of 0.25 degrees.


\begin{table}[]
    \centering
    \begin{tabular}{llr} \toprule
    No. & Variable                          &  Notation   \\ \toprule
    1. & air temperature* (at 2m)           &  $x_{ijt}^{(t2m)}$ \\
    2. & sea surface temperature            &  $x_{ijt}^{(sst)}$ \\
    3. & 10 metre u-component of wind       &  $x_{ijt}^{(u10)}$ \\
    4. & 10 metre v-component of wind       &  $x_{ijt}^{(v10)}$ \\
    5. & 100 metre u-component of wind      &  $x_{ijt}^{(u100)}$ \\
    6. & 100 metre v-component of wind      &  $x_{ijt}^{(v100)}$ \\
    7. & mean sea level pressure            &  $x_{ijt}^{(msl)}$ \\
    8. & volumetric soil water layer 1      &  $x_{ijt}^{(swvl1)}$ \\
    9. & geopotential pressure level on 500 hPa     &  $x_{ijt}^{(geo500)}$ \\ \bottomrule
    \end{tabular}
    \caption{Meteorological variables (data) used in the study. *=not only input variable but output variable as well. A whole dataset is denoted with $x_{ijt}^{(k)}$, where $k$ represents the arbitrary data variable. The true output is represented as $y_t$, prediction output as $\hat{y}_t$.}
    \label{tab:vars}
\end{table}


These meteorological data were then further treated for data reduction by temporal averaging. Downsampling was performed for each meteorological data variable separately, on a fortnight (semi-monthly) basis. This means that the original hourly data were transformed into averaged fortnight data. Hence, two data samples were created for each month, the first sample describing the observations in the first fortnight of a given month and the second sample for the second one. This way, 24 downsampled climate data samples per year were generated. A custom notation of describing of the semi-monthly data was utilised in this paper, i.e., the $\tau_1$ represents the first fortnight of a given month, and $\tau_2$ represents the second fortnight.

The spatial treatment of the data was carried out as follows: First, these 9 different meteorological variables were considered and visualised using the coordinate (geographical) plots. The ERA5 variables were obtained in a regular grid, consisted of a very large sized area incorporating almost the whole Europe including Iceland, part of the northern Africa and almost a half of the Atlantic towards the USA. Three specific problems arose with the incorporation of such amount of data, e.g., (1) It was extremely difficult to process the complete available area with all available meteorological variables due to computational limitations; (2) It seemed intuitive that filtered and concrete subsets of data should lead to better DL performance than tons of unfiltered data; (3) Specific predictor variables, such as the $x_{ijt}^{(sst)}$, were only available at certain areas, i.e., the sea, while for land areas these values were not defined, which could be problematic for the prediction stage. These problems suggest that subsets of data need to be selected before further modelling. We called these subsets as ``geographic area selection'' (GAS), and their purpose is to obtain relevant geographic areas for predicting the outputs ($\hat{y}_t$, compared to the true outputs $y_t$) in a given area of study, i.e., Paris and Córdoba in this case. GAS were obtained for Paris and Córdoba by calculating the Pearson's correlation coefficients for each meteorological variable for each geographic coordinate available. Then, rectangular images considering the most relevant areas were selected to form images, for each predictor variable (geographic areas for each predictor variable were allowed to be different). Image sizes of $33 \times 33$ were empirically recognised as a compromise between the geographical area coverage on one hand and a homogeneity of the correlated areas on the other (larger image sizes would expose areas with less homogeneous values of correlation coefficients, smaller images would omit relevant geographical information).

Pearson's correlation coefficients between each predictor variable and a temperature in Paris or Córdoba ($y_t$) were calculated as follows:
\begin{equation}
    corr_{x_{ij}^{(k)}} = \rho \left (x_{ijt}^{(k)}, y_{t'} \right ),
\end{equation}
where $\rho$ denotes the correlation coefficient calculation, $x$ denotes one of the 9 available data variables, indices $i,j$ denote the pair of location coordinates (latitude, longitude) and $t$ is a time index. $t'$ represents the delayed time index and is used in combination with a variable $y$ that represents the given area temperature at time $t$, $t-1$ ($\tau_1$) or $t-2$ ($\tau_2$). Variable $k$ represents the given predictor (explanatory) variable. It must hold that $lat_{min}^{(k)}<i<lat_{max}^{(k)}$ and $long_{min}^{(k)}<j<long_{max}^{(k)}$, where $lat_{min}^{(k)}, lat_{max}^{(k)}, long_{min}^{(k)}, long_{max}^{(k)}$ define the GAS area. In the next subsections, the two GAS procedures carried, i.e., one for Paris and the other for Córdoba will be presented in detail.

\subsection{Geographic Area Selection for Paris}

In order to obtain the GAS, correlation analyses for each variable are performed between the averaged climate data predictors and an averaged temperature in a given study area (city), considering a possible synoptic relation between predictive and target variables. They are performed specifically for each predictor to obtain the most correlated areas with temperature in a given city. Since forecasts are always made in advance, both the coincident (present) and time-delayed (past) scenarios of correlation analyses are looked for, and a compromise between the two is taken when selecting the GAS. 

\begin{figure}
    \centering
    \includegraphics[width=\textwidth]{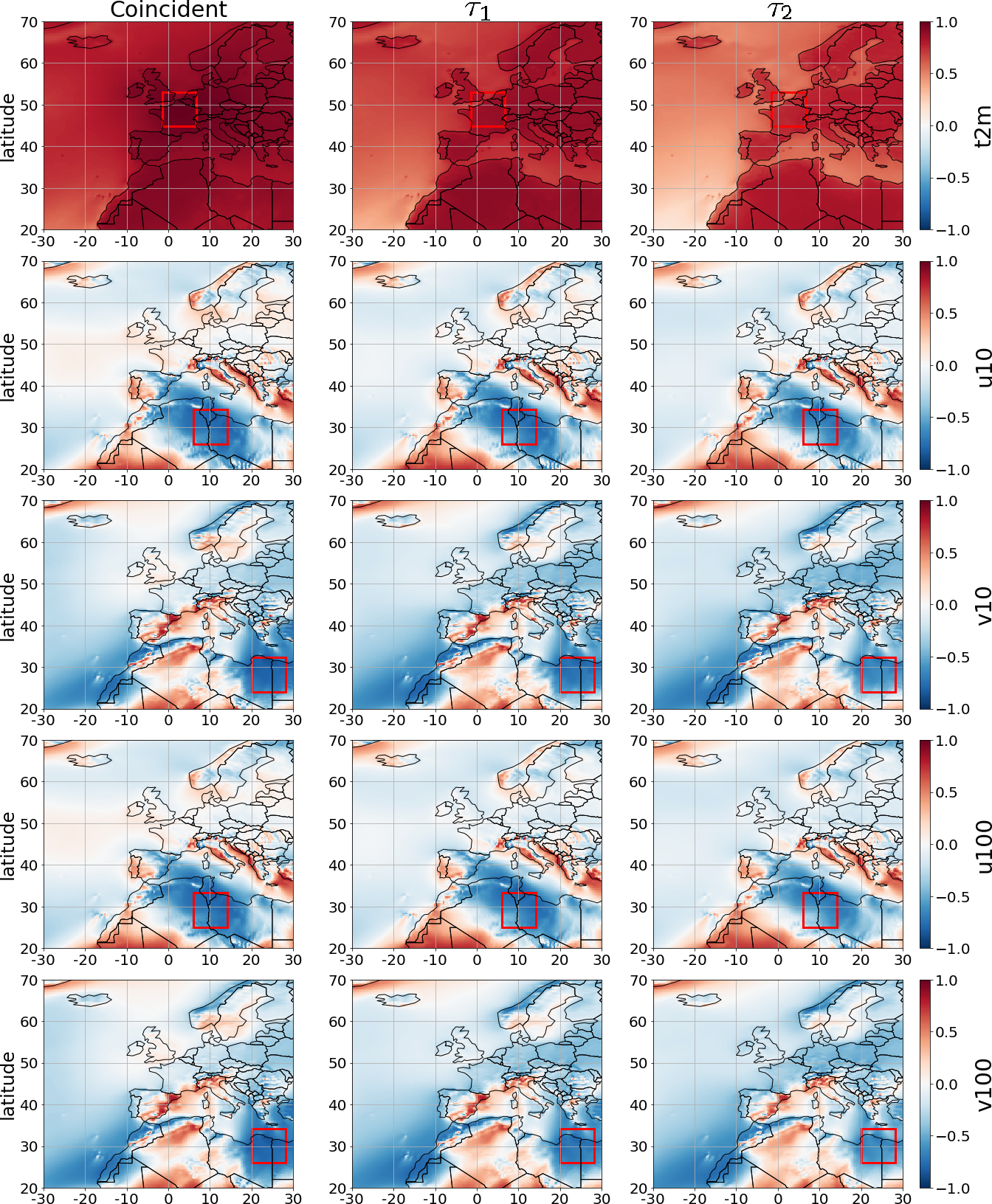}
    \caption{Correlation analysis (Paris) first part of the variables. The three columns represent the Pearson's correlation analyses between the $y_{t}$ in Paris and the each geographic coordinate for each variable $x_{ijt}^{(k)}$. "Coincident"=Pearson's correlation coefficients between the coincident pairs; "$\tau_1$"=Pearson's correlation coefficients between pairs delayed for $\tau_1$, "$\tau_2$"=Pearson's correlation coefficients between the pairs delayed for $\tau_2$. The red rectangles inside the figures denote the regions with highest or lowest correlation coefficients.}
    \label{fig:paris_1_corr}
\end{figure}

Coincident scenario considers time--coincidental pairs of the temperature in Paris (target) and a given predictor variable for each geographic coordinate, e.g., the series of a predictor variable for a given geographic coordinate $x_{ij}$ from Jan-$\tau_1$'1950 to Dec-$\tau_2$'2021 and the $y_{t}$'s in Paris from Jan-$\tau_1$'1950 to Dec-$\tau_2$'2021. The $\tau_1$ time delay scenario depicts the Pearson's correlation analysis between the time delayed $y_{t}$ in Paris for $\tau_1$, e.g. the series of given predictor variable for given geographic coordinate $x_{ij}$ from Jan-$\tau_1$'1950 to Dec-$\tau_1$'2021 and the $y_{t}$'s in Paris from Jan-$\tau_2$'1950 to Dec-$\tau_2$'2021 (note that the sample size decreases for 1 instance in this case). In turn, the $\tau_2$ time delay scenario depicts the Pearson's correlation analysis between the time delayed $y_{t}$ in Paris for $\tau_2$, e.g. the series of given predictor variable for given geographic coordinate $x_{ij}$ from Jan-$\tau_1$'1950 to Nov-$\tau_2$'2021 and the $y_{t}$'s in Paris from Feb-$\tau_1$'1950 to Dec-$\tau_2$'2021 (note that the sample size decreases for 2 instances in this case). Results are visualised onto a geographic map and are interpreted with the help of a colour-bar, where the darker the red or blue colour symbolises the larger the positive or negative correlation, respectively. Figure \ref{fig:paris_1_corr} depicts the three Pearson's correlation analyses for Paris.

\begin{figure}
    \centering
    \includegraphics[width=\textwidth]{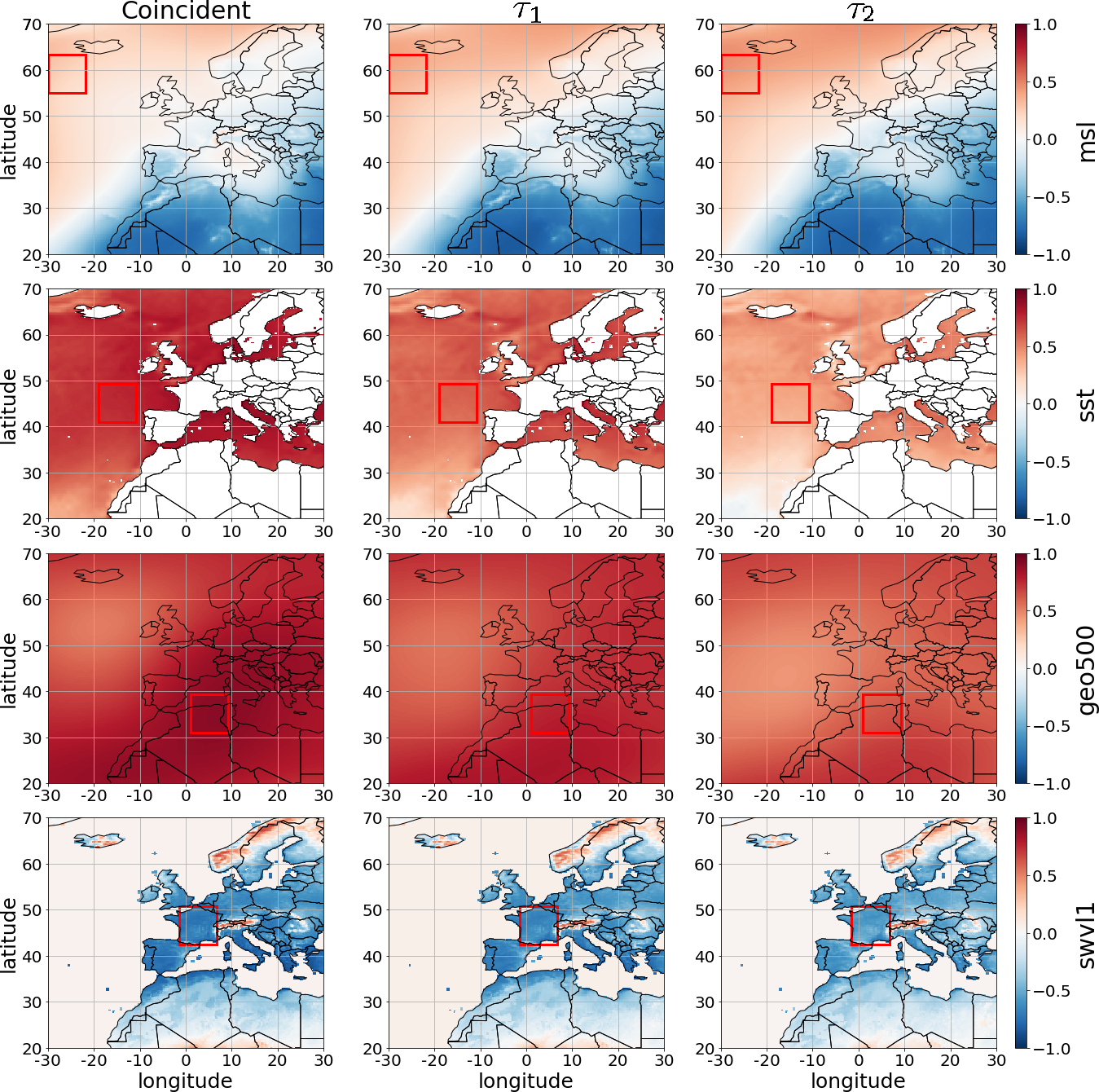}
    \caption{Correlation analysis (Paris) second part of the variables. The three columns represent the Pearson's correlation analyses between the $y_{t}$ in Paris and the each geographic coordinate for each variable $x_{ijt}^{(k)}$. "Coincident"=Pearson's correlation coefficients between the coincident pairs; "$\tau_1$"=Pearson's correlation coefficients between pairs delayed for $\tau_1$, "$\tau_2$"=Pearson's correlation coefficients between the pairs delayed for $\tau_2$. The red rectangles inside the figures denote the regions with highest or lowest correlation coefficients.}
    \label{fig:paris_2_corr}
\end{figure}

As expected, the $y_{t}$ data variable is the most correlated to itself among all predictor variables. Regions near Paris score correlation coefficients near +1, the further we go, the lower the correlation coefficients are. Land is more correlated than the sea: over the Atlantic correlation coefficients in average score values around +0.5. The further we go to the north or south, the lower the correlation coefficient. Similar latitudes on the other hand maintain similar correlation coefficient values. The $\tau_1$ delay is as expected a bit less correlated and the $\tau_2$ delay even less. For the latter, latitudes below $30^\circ$N score correlation coefficients near zero, therefore they cannot deliver much of an information value towards predictions of the $y_{t}$ in Paris. In this case, the GAS region is centred in Paris and extends symmetrically to north, south, east and west.

Next, we analyze the pair of wind components at 10 meters. The $(x_{ijt}^{(v10)}, y_{t})$ pair obtains higher levels of correlation coefficients than the $(x_{ijt}^{(u10)}, y_{t})$ pair. $(x_{ijt}^{(v10)}, y_{t} )$ pair is similarly correlated considering the $\tau_1$ or $\tau_2$ delay with the coincident, and can thus be treated as a stable (or even leading) indicator, although of lower correlation magnitudes, approximately -0.25. South Mediterranean area and the north-west (as well as north-east) of Africa score much higher correlation coefficients and the relation even gets stronger by prolonging the delay. Some parts of the eastern and western Europe seem to be highly positively correlated, but the effect is not much homogeneous and the information value towards predicting the $y_{t}$ is questionable.

No significant correlation is found in the central European area for the pair $(x_{ijt}^{(u10)}, y_{t})$ but strong negative correlation is found between $y_{t}$ in Paris and the $x_{ijt}^{(u10)}$ in Mediterranean sea, meaning that the stronger the u-component of wind (westerlies) the lower the Paris temperature. According to the dark blue colour, wind in Mediterranean's should be a good fit. As expected, the 100 meter wind is more homogeneous than the 10 meter wind. Also, the magnitudes of correlation coefficients are maintained even if increasing the delay. Mediterranean's and the north of Africa again play an important role, just like the north of the Norway. Both of the GAS regions were selected at the north of the Africa, $x_{ijt}^{(u10)}$ at latitudes closer to a Greenwich meridian, $x_{ijt}^{(v10)}$ further away.

Both the $x_{ijt}^{(v10)}$ and $x_{ijt}^{(v100)}$ components that represent the north-south component exhibit a semi-homogeneous tunnel located at the north Africa which, we suppose, symbolises the Sirocco wind pattern. Western and eastern parts of the north of the Africa are of strong negative correlations but the latitudes close to the Greenwich meridian are more to the red. Further, the $x_{ijt}^{(u10)}$ and $x_{ijt}^{(u100)}$ components in central Africa are positive as here the easterlies prevail. The GAS regions for $x_{ijt}^{(u100)}$ and $x_{ijt}^{(v100)}$ are as well set at the north of the Africa.

The $x_{ijt}^{(msl)}$ data variable is among the more important variables. It is utterly homogeneous, with two different zones, north-west and south-east. The former (Atlantic) is positively correlated, while the latter (the north of Africa) negatively correlated, meaning that the higher the $x_{ijt}^{(msl)}$ on Atlantic the higher the Paris temperature, and the higher the $x_{ijt}^{(msl)}$ in Africa, the lower the Paris temperature. Both the positive and negative correlations enlarge by extending the delay, thus the $x_{ijt}^{(msl)}$ should be treated as an excellent leading indicator. The GAS is set at the extreme western part of the Atlantic available, at high latitudes, close to Iceland.

The $x_{ijt}^{(sst)}$ data variable is available only for the sea areas, e.g. Atlantic and Mediterranean's. The relation is highly robust for coincident scenario, but drops substantially with the introduction of delay. Most correlated seem to be areas of similar geographic latitudes. On the other hand, the $( x_{ijt}^{(geo500)}, y_{t} )$ pair exhibits much higher relation strengths. Except a central area on the Atlantic which seems to exhibit lower correlation pairs, correlations are above +0.5 and persistent at the delays. The higher the $x_{ijt}^{(sst)}$ or $x_{ijt}^{(geo500)}$ values, the higher the temperature in Paris.

Finally, the $(x_{ijt}^{(swvl1)}, y_{t})$ is addressed, here the prolongation of delay reduces the correlation fit. Strongest negative fit is found to be on the land near across the whole Europe, meaning that the higher the volumetric soil water layer, the lower the Paris temperature. No correlation is found between the $( x_{ijt}^{(swvl1)}, y_{t} )$ pair on sea (either the Atlantic or Mediterranean). The correlation fit worsens by considering northern or southern latitudes, such as north of Africa or Norway, Sweden, Iceland. An outlier, i.e. the Alps, is spotted: here, the Alps are positively correlated. The GAS region is selected to be covering as much as France as possible.

\subsection{Geographic Area Selection for Córdoba}

The results show in this case that the $y_{t}$ data variable is among the most correlated variables again, especially for regions of the central Europe and the north-east of the Africa. Unfortunately, the correlation coefficients quickly reduce by introducing the $\tau_1$ and $\tau_2$ delays. Again, the $y_{t}$ on the land is more correlated than the $y_{t}$ on the sea, e.g., Atlantic; low correlations are found for $y_{t}$ on the sea below latitudes $30^\circ$N. The GAS region is centred at Córdoba.

Correlations for pairs $(x_{ijt}^{(u10)}, y_{t})$ and $(x_{ijt}^{(u100)}, y_{t})$ wind components are similar between themselves. The Mediterranean sea is highly negatively correlated with the $y_{t}$ in Córdoba and the south of the Europe is highly positively correlated, but the fit is not as homogeneous as Mediterranean's. Additionally, the north of the Sahara seems to be positively correlated with the Córdoba's $y_{t}$, but the fit is due to limiting coordinates not homogeneous as desired. Both the v-components seem to exhibit similarly positive correlations behaviour as in the Paris case, with the significant but leaky tunnel between the north of the Africa and south of France. Again, the south-east of the Europe is most negatively correlated. The GAS regions for $x_{ijt}^{(u10)}, x_{ijt}^{(u100)}, x_{ijt}^{(v10)}, x_{ijt}^{(v100)}$ are centred identically to the Paris case.

The $x_{ijt}^{(msl)}$ shows a similar structure to the Paris case, yet very interesting realisation -- the larger the time delay, the higher the correlation. Atlantic and the north-east of the Europe is positively correlated with the Córdoba's $y_{t}$ and the south-east of Europe and north of Africa is again negatively correlated. A similar realisation is with the $x_{ijt}^{(sst)}$ -- the higher the latitude, the higher the correlation coefficient; although the sea surface temperature correlation coefficient gets weaker by increasing the extending the time delay. The $(x_{ijt}^{(geo500)}, y_{t})$ again realises a central Atlantic part less related with the Córdoba $y_{t}$, but for the rest of the regions, especially the northern Africa, it exhibits a highly positively related connection. The $(x_{ijt}^{(swvl1)}, y_{t})$ pair shows a strong negative correlation fit over the land, but weak or none fit on the sea. The strongest negative fit is spotted for the similar latitudes as the Córdoba itself. The GAS for $x_{ijt}^{(sst)}, x_{ijt}^{(geo500)}$ predictor variables are centred identically as in Paris case, while the $x_{ijt}^{(swvl1)}$ is centred to cover the Iberian Peninsula.

\begin{figure}
    \centering
    \includegraphics[width=\textwidth]{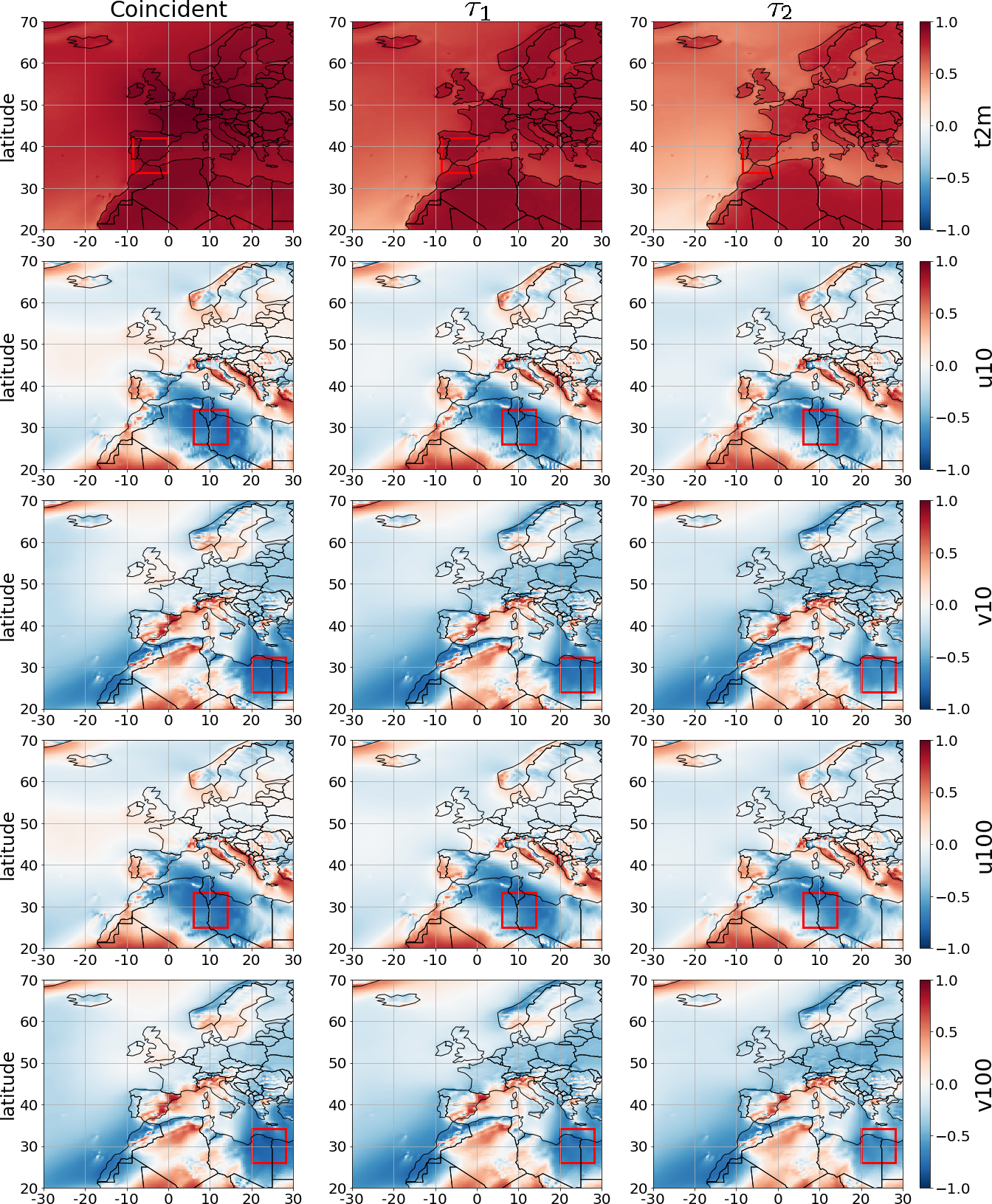}
    \caption{Correlation analysis (Córdoba) first part of the variables. The three columns represent the Pearson's correlation analyses between the $y_{t}$ in Córdoba and the each geographic coordinate for each variable $x_{ijt}^{(k)}$. "Coincident"=Pearson's correlation coefficients between the coincident pairs; "$\tau_1$"=Pearson's correlation coefficients between pairs delayed for $\tau_1$, "$\tau_2$"=Pearson's correlation coefficients between the pairs delayed for $\tau_2$. The red rectangles inside the figures denote the regions with highest or lowest correlation coefficients.}
    \label{fig:cordoba_1_corr}
\end{figure}

\begin{figure}
    \centering
    \includegraphics[width=\textwidth]{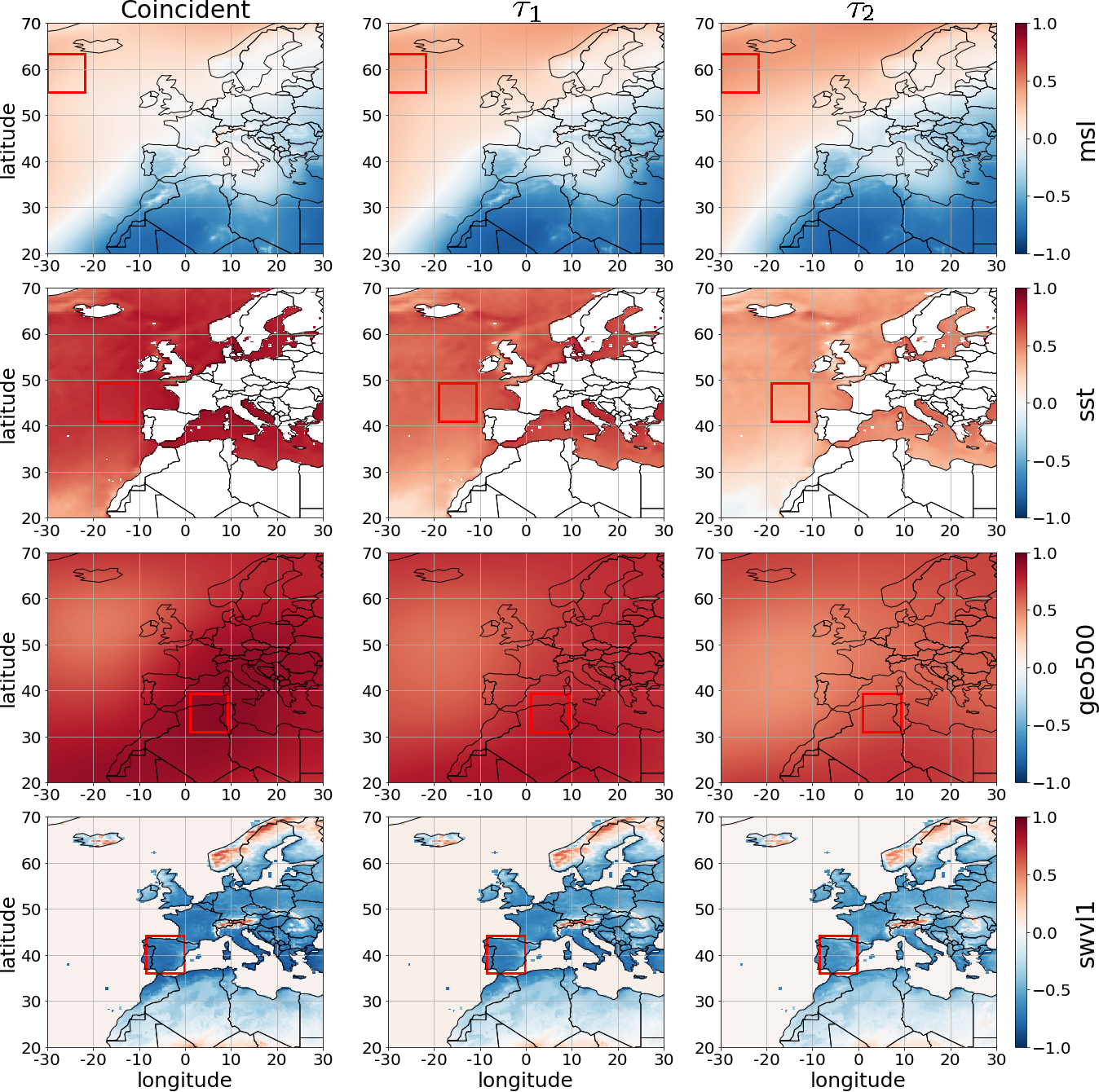}
    \caption{Correlation analysis (Córdoba) second part of the variables. The three columns represent the Pearson's correlation analyses between the $y_{t}$ in Córdoba and the each geographic coordinate for each variable $x_{ijt}^{(k)}$. "Coincident"=Pearson's correlation coefficients between the coincident pairs; "$\tau_1$"=Pearson's correlation coefficients between pairs delayed for $\tau_1$, "$\tau_2$"=Pearson's correlation coefficients between the pairs delayed for $\tau_2$. The red rectangles inside the figures denote the regions with highest or lowest correlation coefficients.}
    \label{fig:cordoba_2_corr}
\end{figure}

The predictive climate variable have been then limited to the GAS regions in order to carry out the air temperature prediction with the ML and DL approaches. Since we are particularly interested in forecasting the summer temperatures, GAS regions were further downsampled to include months from April to August only. Between these, 8 time samples from April to July were considered as input data, and 2 time samples as output model data, i.e., August $\tau_1$ or $\tau_2$. The next subsection describes the further data adjustment procedure.

\subsection{Data adjustment procedure}
The GAS procedure leaved us with the original (unit) data, thus some data adjustment were needed before employing the proposed modelling with AI techniques. Input and output data were first normalised separately. First, the input data ($x_{ijt}^{(k)}$) normalisation within the range $[0,1]$, was employed using the following transformation:
\begin{equation}
    x_{ijt}^{(k)'} = \frac{x_{ijt}^{(k)} - \min_t x_{ijt}^{(k)}}{\max_t x_{ijt}^{(k)} - \min_t x_{ijt}^{(k)}},
\end{equation}
where the $x_{ijt}^{(k)}$ represents the original input set of data from April--July, $x_{ijt}^{(k)'}$ the normalised input climate variables, $i$ and $j$ represent the coordinates (longitude, latitude), $k$ represents each of the nine of the climate variables considered and $t ~\in~ [1950,1951,\ldots,2021]$ represents the time. As it can be seen in the transformation with the index $t$, input data normalisation was performed specifically for each year (still, all the time samples from April--July within a given year were normalised using the same factors). 

The output data ($y$), which effectively represents the given area temperature data in August, either the first or second fortnight, is adjusted twofold. First, it is adjusted using the input data $x$ normalisation factors, as follows:
\begin{equation}
    y_{ijt}' = \frac{y_{ijt} - \min_t x_{ijt}^{(t2m)}}{\max_t x_{ijt}^{(t2m)} - \min_t x_{ijt}^{(t2m)}},
\end{equation}
where $y_{ijt}$ illustrates the original regional temperature output and the $y_{ijt}'$ is the scaled regional temperature output data. However, note that this adjustment does not enssure the normalised data within the range $[0,1]$, rather aggregated numbers close to 1 with a very small variance. After, the adjusted $y_{ijt}'$ is normalised to ensure the $[0,1]$ range as follows (and hence maximise the output variance):
\begin{equation}
    y_{ijt}'' = \frac{y_{ijt}' - \min_t y_{ijt}'}{\max_t y_{ijt}' - \min_t y_{ijt}'}.
\end{equation}
Also, the output data was given as an image (with appropriate $i$ and $j$ coordinates), but by selecting a single pixel (e.g., $i=1,j=1$) only a specific geographical location could be extracted, i.e., $y_{t}''$.

\subsection{Exhaustive feature search}

There are nine different predictors (input variables) included in the analysis carried out. Some of them may appear as redundant (especially regarding the wind) and thus they are suspicious to lower the forecasting skills of the prediction model. Exhaustive Feature Search (EFS) is therefore employed, due to low number of existing predictors, to test all possible combinations of predictors ($2^9=512$). Upon, the best obtained combination is taken for forecasting.

In addition, note that we consider nine different modelling methods applied in the paper (LR, Lasso, Poly, AdaBoost, DT, RF, CNN, RP+CNN and RP+CNN+BIN). Different modelling methods incorporate different training skills, so in order to maximise the training skill (and consequently the forecasting skill), the most suitable combination of predictors is sought for each modelling method specifically. The best combination of predictors may thus differ between methods (no universal solution may work equally well for all methods). EFS is conducted by first generating a list of all possible combinations of predictors. Next, each of the possible combination of predictors is trained for each model and forecasts are run. The mean squared error ($mse$) of the forecasts is then calculated compared to the true values. Finally, the combination of predictors with lowest $mse$ (for each modelling method specifically) is taken as a best combination of predictors. The best set of features for each prediction model will be shown in the results section.

\section{Proposed computational frameworks based on AI for long-term temperature prediction}\label{sec:methods}
This section presents the three proposed computational frameworks for long-term air temperature prediction. All the proposed computational frameworks exploited the same data as described above. However, slight further modifications and adjustments were applied to adjust the data to each of the methods unambiguously. Especially, the data sequencing procedure, which will be explained for each method distinctively, provided large differences regarding the data exploitation among the three frameworks.

\subsection{Computational framework 1: Convolutional Neural Networks}

CNNs are universal, deep learning networks for processing images and videos, either for regression, classification, segmentation or identification purposes \cite{dhillon2020convolutional}. The hearth of the CNNs is the CNN kernel, a matrix or tensor with trainable weights. Weights are typically randomly initialised and are adjusted during the CNN training. CNN kernel with weights performs a mathematical operation of convolution and produces CNN's hidden layers, so-called feature map. Within a single hidden layer many feature maps are typically produced by many distinctive CNN kernels. Feature maps are usually of lowered dimensionality compared to the inputs. Such dimensionality reduction depends on the CNN kernel size and is usually minimal. Rather, dimensionality of feature maps is controlled by the pooling operation. Pooling only adjusts the dimensionality, but does not provide any trainable weights. Several pooling strategies, such as maximum or average pooling, exist. Most often, pooling is used in conjunction with a convolution layer, in a stacked architecture where convolution is first applied and then pooling follows. In a multi-layered CNN such stacking combinations are applied several times, meaning that the dimensions and number of feature maps may change several times between CNN input and output. The CNN output also is a feature map. It can be either treated as an image or alternatively be flattened into a single regression value or classification probability using a dense layer. For complex regression or classification problems, several dense layers can be applied.

\begin{figure}
    \centering
    \includegraphics[width=.8\textwidth]{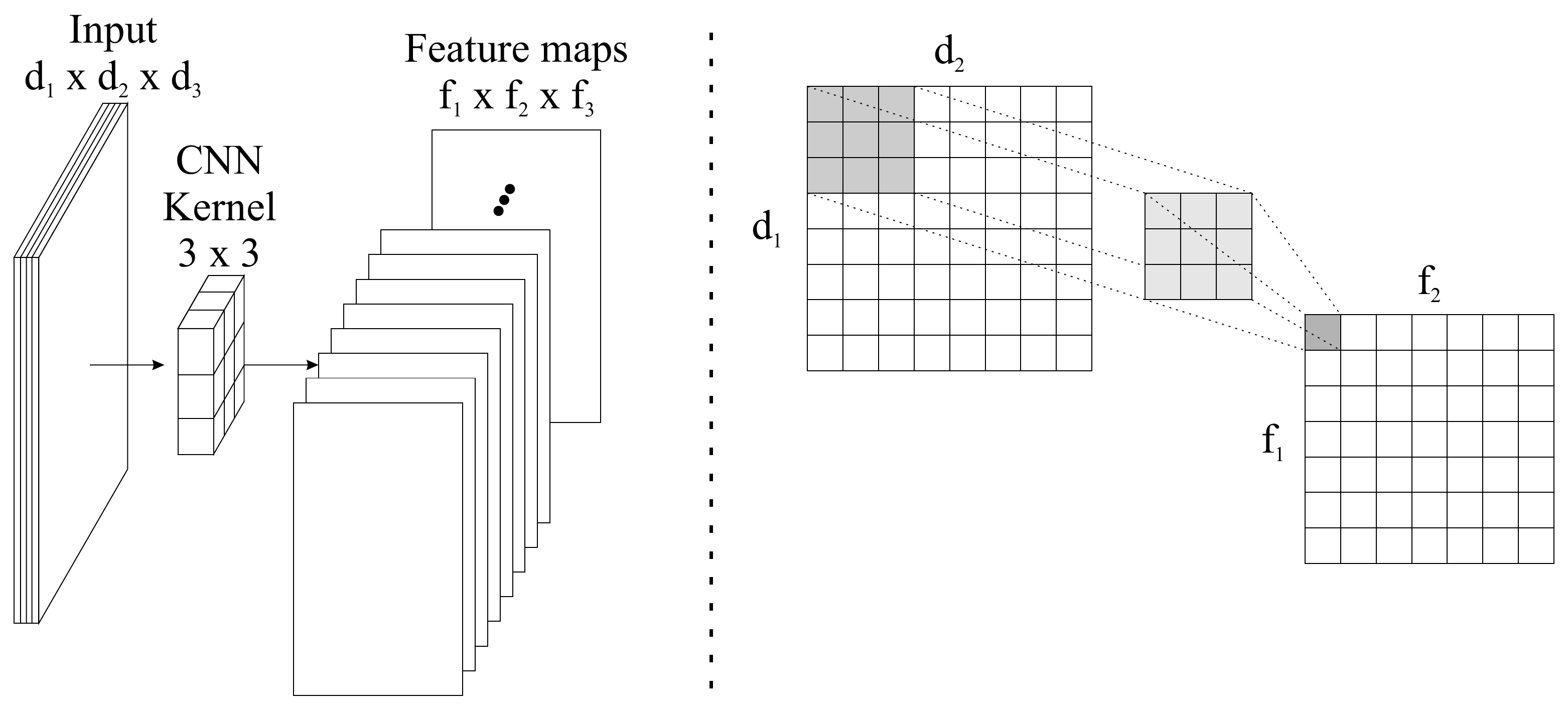}
    \caption{Demonstration of a CNN convolution.}
    \label{fig:gen_cnn}
\end{figure}

Figure \ref{fig:gen_cnn} shows a general example of the CNN convolution. Figure is divided into two schemes representing the same concept. The left scheme is more abstract, while the right more detailed. Input into the demonstrated CNN is the image of dimensions $d_1 \times d_2$ and $d_3$ channels (images typically incorporate three channels, the red, green and blue). Input image is convolved with the demonstrated CNN kernel of dimensions $k_1 \times k_2$, where $k_1=k_2=3$. Procedure of convolution is repeated $f_3$-times, each time with distinctively initialised kernel weights. In such a way, $f_3$ feature maps of dimensions $f_1 \times f_2 \times f_3$ are generated. The detailed scheme represents the extraction of dark grey coloured subimage to be convolved with the light grey coloured kernel. Typical CNN convolution multiplies the values of subimage with the kernel weights element-wised and sums them. After, the bias $b$ is added. Finally, the result is saved as a single component into the feature map (on figure represented by the dark greyed colour single pixel). This process is repeated by gradually moving the dark grey coloured subimage over the rest of the image, a process thoroughly controlled by the kernel stride parameters. Both the overlapping or non-overlapping scenarios of subimages may be applied. After all suitable combinations, given by image size, kernel size and kernel stride, are gone through, a complete feature map is built.

\begin{figure}
    \centering
    \includegraphics[width=\textwidth]{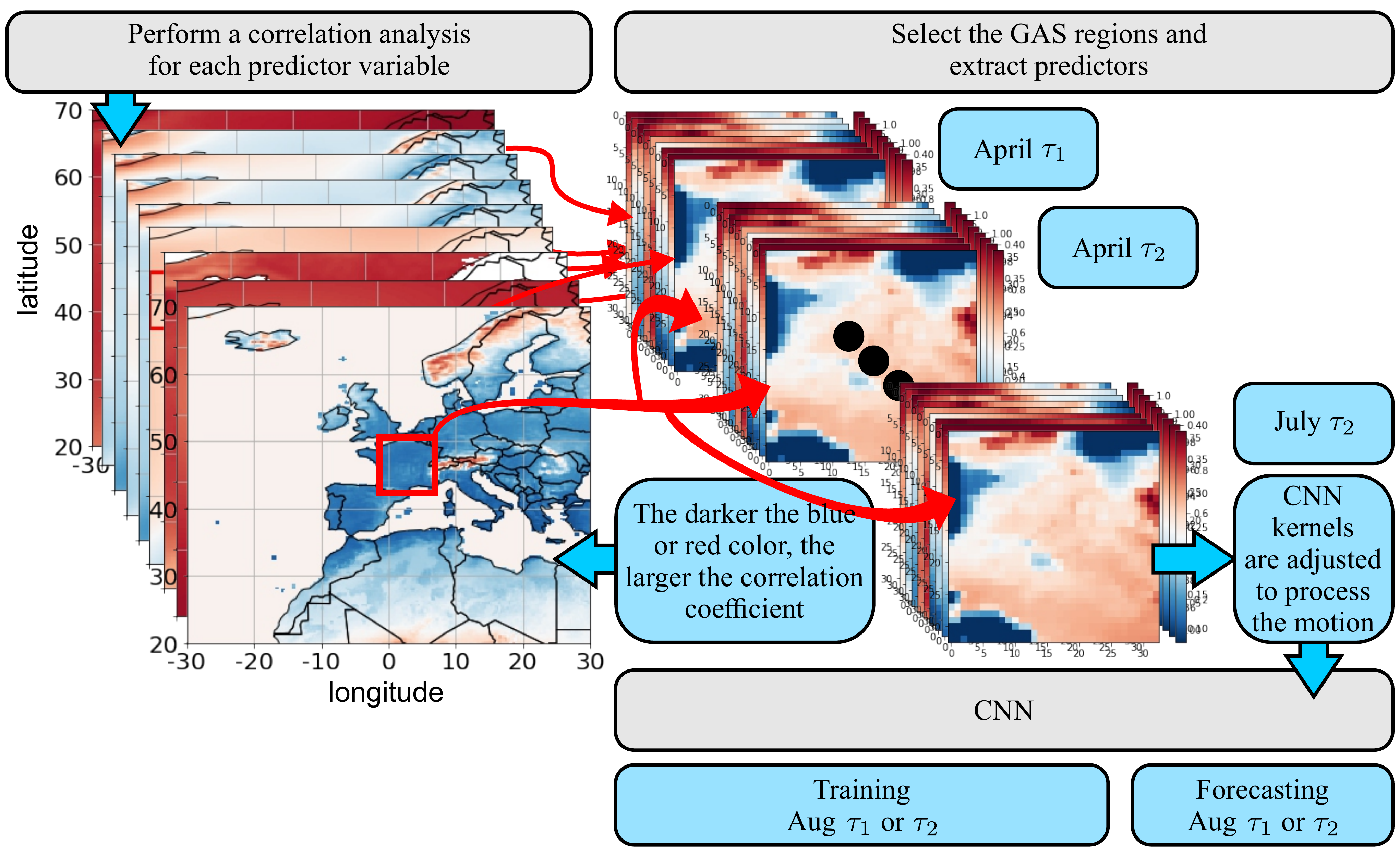}
    \caption{The three grey rectangles represent the proposed workflow. The correlation analysis is used to derive the GAS regions. Next the data sequencing follows to build the instances. Finally, the supervised training with out-of-sample forecasts is employed.}
    \label{fig:cnn_use_case}
\end{figure}

Following the introductory demonstration of CNNs, the CNN computational framework as used in the study is presented. The use-case diagram of the originally proposed CNN computational framework can be visualised in figure \ref{fig:cnn_use_case}. Figure is organised as a flowchart, and addresses three important steps of CNN exploitation (each of these steps is indicated by a grey coloured rectangular box). First, a correlation analyses of the fused data are run as shown in section \ref{data}. Correlation analyses provide the GAS regions, one per predictor, which are in the figure exhibited by red symmetrical rectangles. Positions of GAS regions are fixed for a given variable but may differentiate between variables.

Next, the data sequencing step follows. The purpose of the data sequencing is to build a multivariate data structure similar to the moving images (video). There are 9 predictors, each of them forms a single channel. Predictor values are taken from GAS regions for two of the each monthly fortnights ($\tau_1,\tau_2$). Months from April to July are covered, meaning that 8 different images that form a motion with a sequence length 8, are introduced. Processing of the images is always from the oldest to latest, as depicted on the figure \ref{fig:cnn_use_case}. First comes the April's $\tau_1$, followed by April's $\tau_2$,..., the last image is July's $\tau_2$. The whole motion of images is called an instance. Each instance represents an individual year, there will be so many instances as is the number of years of data available. However, not only input variables undergo slight data modifications but also the output variable does. The CNN output $\hat{y}_{ijt}''$ is organised as an image and needs to be compared with an image during the CNN training to derive the weight corrections. Namely, if the dimension of the CNN output is lower than the CNN input (due to CNN convolutions), the true output image $y_{ijt}''$ dimensionality needs to be thus lowered as well. A symmetric lowering of dimensions is employed as $y_{ijt}''' = y_{ijt}'' [l:n-l, l:n-l]$, where $l$ controls the level of lowering. Due to symmetry, the centre of the so reduced true image is maintained.

\begin{figure}
\centering
\begin{subfigure}{1.0\textwidth}
    \includegraphics[width=\textwidth]{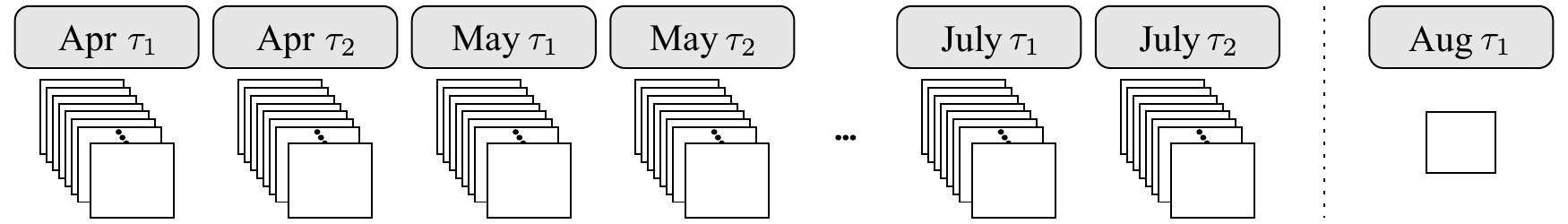}
    \caption{Temperature forecasts for sooner fortnight ($\tau_1$).}
\end{subfigure}
\hfill
\begin{subfigure}{1.0\textwidth}
    \includegraphics[width=\textwidth]{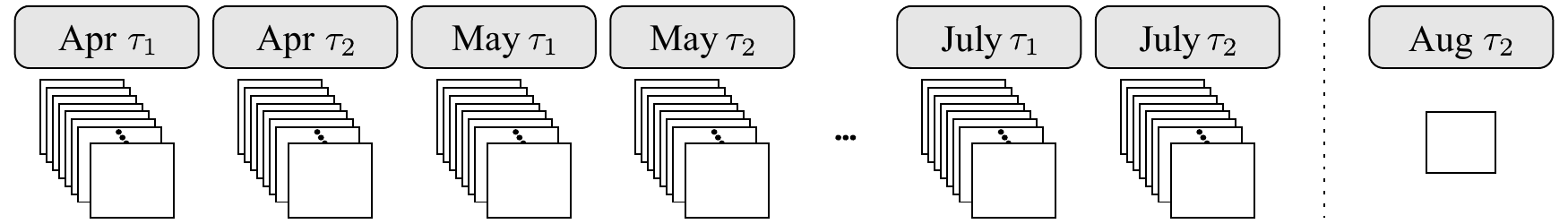}
    \caption{Temperature forecasts for later fortnight ($\tau_2$).}
\end{subfigure}
\caption{Forecast diagram. One CNN input instance consists of 4 consecutive months from April to July, two fortnights per month. An output is organised as either Aug $\tau_1$ or Aug $\tau_2$.}
\label{fig:time_horizon}
\end{figure}

Finally, the CNN training and forecasting procedures are run. Instances (72 of them) are divided into two strictly non-overlapping sets, the training and forecasting sets. As output, either the August's $\tau_1$ or $\tau_2$ is applied. Figure \ref{fig:time_horizon} better describes the outline of a single instance and forecast output.

\begin{figure}
    \centering
    \includegraphics[width=\textwidth]{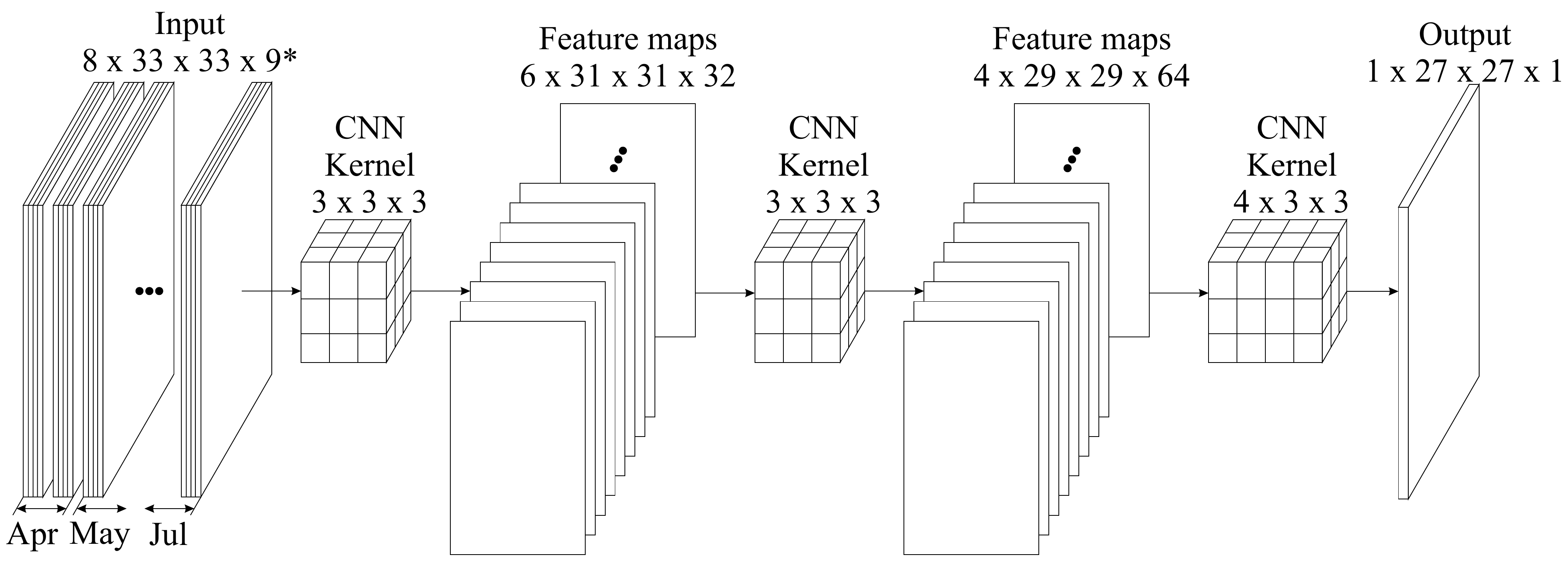}
    \caption{CNN architecture. The input consists of a sequence of images (or a motion). Each image consists of $9^*$ different channels, 2 images per month. Months from April to July are covered within the input data, either August $\tau_1$ or August $\tau_2$ in the output. The CNN processes the input data using 3 separate 3D kernels, hence 2 sets of feature maps are generated (the third set of feature map is the output). The output is organised as a 2D image, with adjusted dimension size.}
    \label{fig:cnn}
\end{figure}

A detailed architecture of proposed CNN computational framework is visualised in Figure \ref{fig:cnn}. The figure is adjusted to exhibit a single instance only. Input size of an instance is $8 \times 33 \times 33 \times 9$ (time samples $\times$ x-axis size $\times$ y-axis size $\times$ number of channels, respectively). A 3D CNN kernel of $3 \times 3 \times 3$ is proposed to create the first layer of feature maps. The 32 feature maps are generated, each of the size $31 \times 31$, and the sequence is lowered from 8 to 6. Next CNN hidden layer is applied by a CNN kernel of $3 \times 3 \times 3$. The number of feature maps is increased to 64 and sequence length decreased to 4. The final CNN kernel is of customised dimensions $4 \times 3 \times 3$ to assure the single-channelled output image. The CNN output image dimension is equal to $27 \times 27$ and that is also the dimension of the true output image dimension, hence $l=(33-27)/2=3$.

\subsection{Computational framework 2: ML methods}

Six different ML methods are also implemented and tested to be applied in the air temperature forecast problems considered. Three of them are deterministic, such as linear regression (LR), Lasso regression (Lasso), and Polynomial regression (Poly), and three of them more sophisticated, such as AdaBoost, Regression Trees (DTs), and Random Forest (RF). In what follows, each ML method is presented briefly.

LR is a traditional, easy-to-use, and a low-complex shallow modelling method. It is able to capture linear, as well as non-linear connections between input and output variables. Due to its versatility and extremely quick processing, LR is one of the more popular benchmark methods among researchers. The original version of the LR, incorporating the ordinary least square minimising algorithm, was proposed by mathematician Gauss \cite{weisberg2005applied}.

Lasso regression is an advanced and automated modelling method that combines the feature selection with original LR methodology. Developed by Tibshirani \cite{tibshirani1996regression}, the motivation of the Lasso is to omit the redundant and less relevant predictors from the model and thus improve the model performance. The objective of the Lasso is to minimise the variance of regression parameters on the behalf of so-called shrinkage parameter $\lambda$. An original objective function (which is also used during the experiments), is stated as $1/2||Y-X\beta||^2_2+\lambda||\beta||_1$, where the term $Y-X\beta$ represents the residual sum of squares and the mathematical notation $||$ represents the norm. The Lasso can be also seen as a regularisation method and is especially suitable for datasets with higher number of features and for datasets with higher level of uncertainty.

Polynomial regression works with the traditional LR, but transforms predictors non-linearly before the use. By its nature, it increases number of features, but this increase can be controlled conveniently by the parameter setting on level of degrees.

DTs are a popular shallow estimators which are originally purposed for classification tasks, but are also capable of solving the regression tasks. Pioneered by Quinlan \cite{quinlan1986induction} and Breiman et al. \cite{breiman2017classification}, DTs became a robust and reliable ML estimators and have since inception been applied to a large number of prediction problems, including meteorological applications and climate prediction tasks \cite{geetha2014data,wei2020decision,ngo2021novel}.

AdaBoost (also Adaptive Boosting) was not proposed as a self-standing modelling method \cite{freund1996experiments}. Instead, it bases on one of the other underlying estimators, typically DTs. Purpose of the AdaBoost is to build an ensemble of DTs with different subsamples \cite{schapire2013explaining}. There have been recent successful applications of AdaBoost to prediction problems in climate and related tasks such as \cite{xiao2019short,asadollah2022prediction}.

RFs are a state-of-the-art ensemble classification and regression methods that similarly as AdaBoost use DTs as an underlying estimators \cite{ho1995random, breiman2001random}. RFs also exploit repetitive subsampling to build many weak-learners, which are then managed into a strong-learner using a voting mechanism. Some recent climate applications with RFs are \cite{grazzini2020extreme,grazzini2021extreme,park2016drought}.

All the adopted ML methods are trained and tested using the same data samples. However, some further modifications of ML methods data are required compared to the CNN data, since adopted ML methods cannot process images, nor motions of images. A simple remediation to adjust the data for ML methods is employed. The CNN data structure is taken as a baseline, from which maximum values (a single pixel) for each channel are extracted. Initially, we have tested other extractions, such as minimum, average or median, but extraction of maximum value was realised empirically as the best hit. The process of extracting the maximum value is repeated for each fortnight and the temporal data are stacked horizontally as individual instances. Formally, the extraction of maximums is denoted in Equation \eqref{eq:max}.
\begin{equation}
\label{eq:max}
    x_{t}^{(k)''} = \max_{i \in 1,2,\ldots,n} \left( \max_{j \in 1,2,\ldots,n} \left( x_{ijt}^{(k)''} \right) \right),
\end{equation}
where $x_{t}^{(k)''}$ denotes the ML adjusted data. The equation is saying that the spatial dependencies are removed by picking the maximum point within each channel of the image. Only two indices, $t$ and $k$, which represent time and type of the predictor variable (channel), remain in the ML adjusted data. In this way, a lot of the data is lost. This can be either positive, due to significant filtration of data redundancy, since we assume that the climate data close together are similar. Or, it can also be negative, due to losing many of the details. Next, the use-case diagram of ML methods is demonstrated.

\begin{figure}
    \centering
    \includegraphics[width=.8\textwidth]{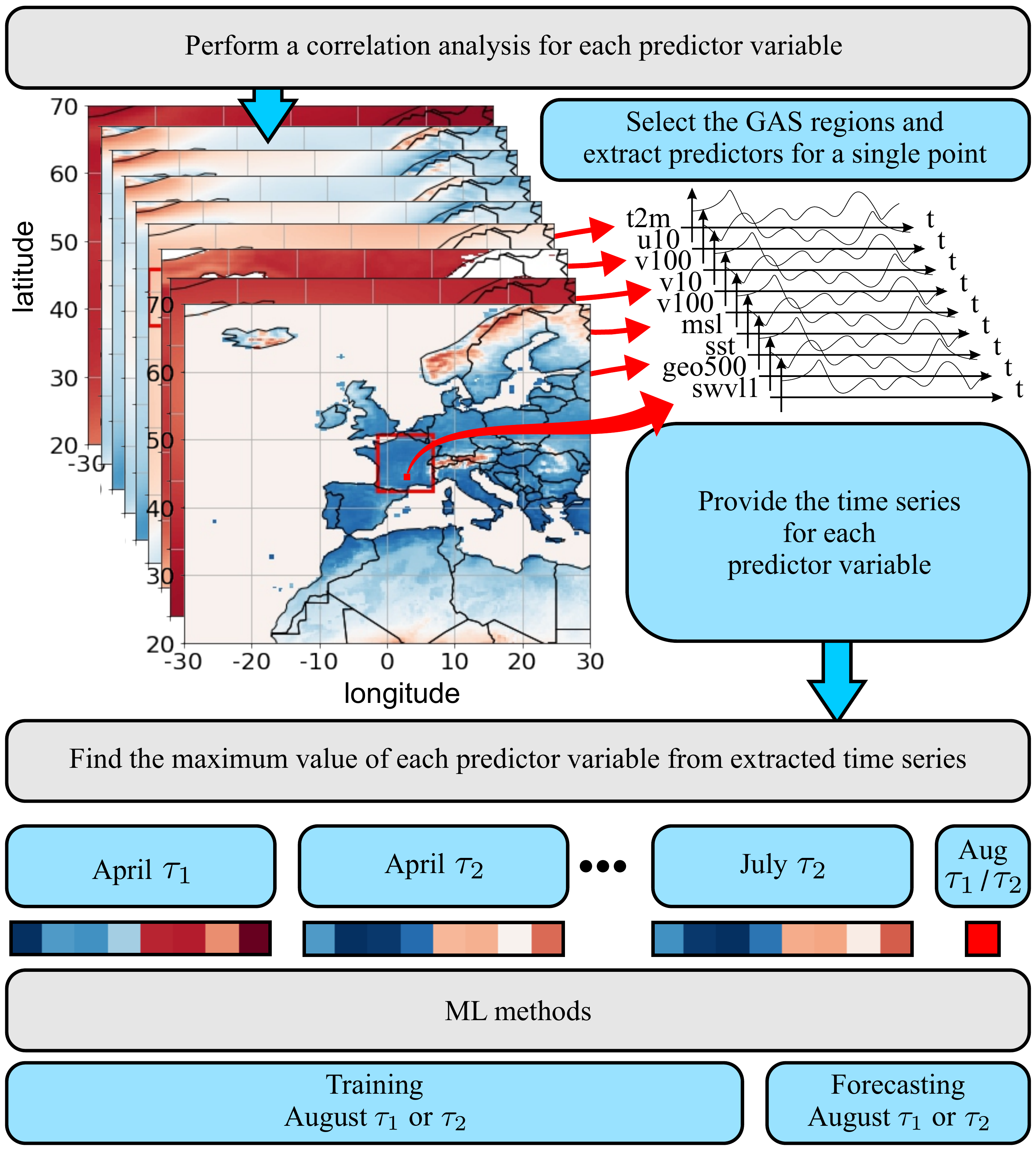}
    \caption{The three grey rectangles represent the proposed workflow. The correlation analysis is used to derive the GAS regions. Next, the data sequencing follows to make the adjustments for ML data. Finally, the supervised training with out-of-sample forecasts is employed.}
    \label{fig:ml}
\end{figure}

The use-case diagram of ML methods is shown in Figure \ref{fig:ml}. ML adjusted data follows the identical correlation analyses as CNN data to obtain GAS regions. Adjusting the ML data by maximising each channel is thus seen as additional adjustment procedure. Maximum values are on the use-case diagram demonstrated with a small but visible red rectangular point and red arrows are driven out of them. The process is repeated for each predictor and individual maximum values are collated to form a vector of $1 \times 9^*$. The process is further repeated for each fortnight (8 in total) and individual vectors of predictors are stacked into an instance to form a vector of $1 \times 9^*x8$, i.e., $1 \times 72^*$.

\begin{figure}
    \centering
    \includegraphics[width=.7\textwidth]{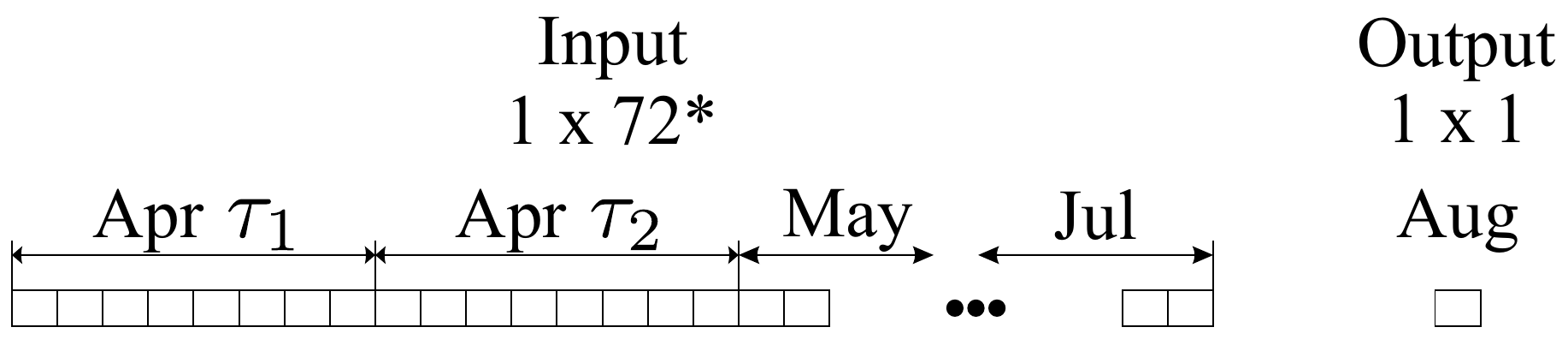}
    \caption{Architecture of ML data. The input consists of a $72^*$ featured vector which represent the sequence of months variables from April to July. The output is organised as a single value for regression task and either represents the August's $y_{t}$ in $\tau_1$ or $\tau_2$ prediction horizons.}
    \label{fig:ml_arch}
\end{figure}

The architecture ML data is visualised in Figure \ref{fig:ml_arch}. Each fortnight is represented with the $9^*$ predictors. The output is organised as a single $1 \times 1$ value.

Due to the ML data adjustments, the ML methods are fed with significantly less data. Theoretically, this is a drawback, since less data carry less information. Although, performed tests have revealed that model performance is not hurt much by incorporating less data. Therefore, we proposed to build a variant of CNN that would operate on the ML adjusted data. ML adjusted data is seen by CNNs as 1D. We proposed to transform the 1D data into images by using the RP to make them more comfortable for CNNs. In this way, the same data were exploited in the transformed way. The next subsection describes the combination of RP and the CNN.

\subsection{Computational framework 3: Recurrence plot with Convolutional neural network}

This subsection represents the theoretical outline of the RP transformation and associated transformation process. Two types of RPs are used, the classic and binarised. Next, the use-case diagram and the used RP+CNN(+BIN) architecture are demonstrated.

\begin{figure}
    \centering
    \includegraphics[width=.8\textwidth]{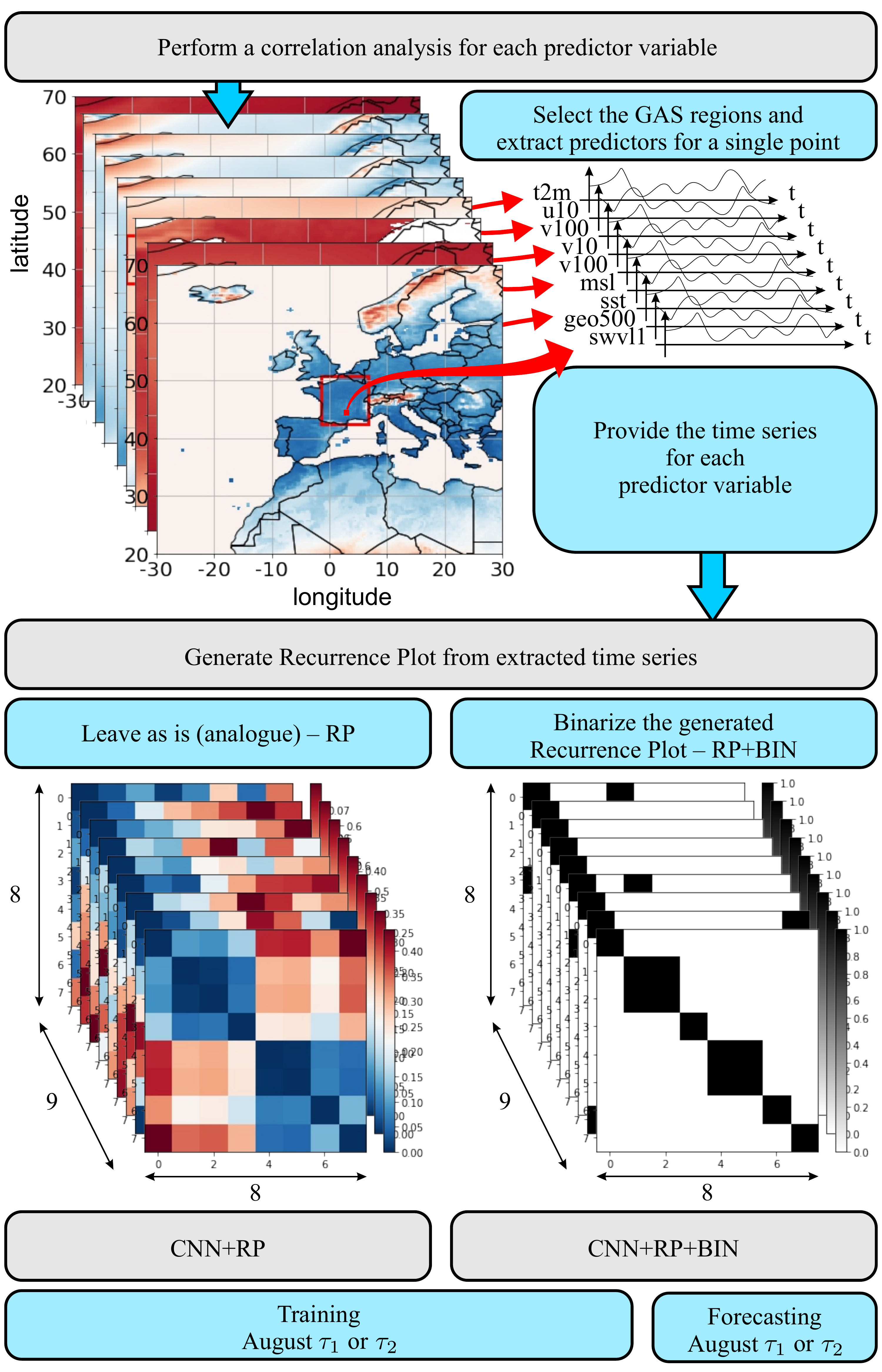}
    \caption{The three grey rectangles represent the proposed workflow. The correlation analysis is used to derive the GAS regions. Next, the data sequencing follows to make the adjustments for ML data. Then, RPs are build. Finally, the supervised training with out-of-sample forecasts is employed.}
    \label{fig:rp_cnn_use_case}
\end{figure}

RP transformation is in general a mathematical process of subtracting and deriving a norm of the two displaced time series elements and the result is graphically visualised as an image. Only a single image with $9^*$ channels is created from ML adjusted data, where the shape (ornament) of an image represents the time series of each channel. Formally, the input vector ($\mathbf{g}_p$) is represented as follows (mathematical expressions are summarised from pyts library  \cite{faouzi2020pyts}, please note that the variable names and indices are customised):
\begin{equation}
    \mathbf{g}_{a} = (g_a, g_{a + \tau}, \ldots, g_{a + (b - 1)\tau}), \quad
        \forall a \in \{1, \ldots, c - (b - 1)\tau \},
\end{equation}
where the $a$ runs from 1 to $c - (b - 1)\tau$. In case $\tau=1$, one can state a simplified representation:
\begin{equation}
    \mathbf{g}_a = (g_1, \ldots, g_c).
\end{equation}
where the $c$ represents the number of timestamps. The RP calculation is derived by accounting for two iterative variables, i.e., $a,d$. The output of the RP is a 2D image and is symmetric over the diagonal, formally,
\begin{equation}
    R_{a, d} = \Theta(\varepsilon - \| \mathbf{g}_a - \mathbf{g}_d \|), \quad
        \forall a,d \in \{1, \ldots, c - (b - 1)\tau \}.
\end{equation}
Mathematical operator $\| \cdot \|$ represents the Euclidean 2D norm between the two timestamps $a$ and $d$, and the $\varepsilon$ represents the so-called threshold. Threshold is optional. If used, the RP image is binarised, if not, the RP image is the left analogue. Both of the options have been tested in this computational framework, the analogue we denote as RP+CNN, the binarised as RP+CNN+BIN. By nature, threshold is one of the tuning parameters. If applied, the RP image undergoes the Heaviside step function, denoted as $\Theta$, which then delivers the binarisation. Three different scenarios of binarising the RPs exist. First of them sets the given percentage of $1-p$ of pixels with lowest values to 0 and the rest of the pixels of percentage $p$ to 1. The second seeks for the maximum value of RP and sets the individual pixels which values are less than given percentage of $1-p$ of the maximum value to 0. Others are set to 1. The third option comes with manual specification of threshold. Figure \ref{fig:rp_cnn_use_case} shows the use-case diagram of the RP+CNN(+BIN) methods. Correlation analyses are identical to CNN and ML computational frameworks. Again, maximal values are derived from the coordinate data and the RPs are generated with this adjusted data. The analogue (originally obtained) RP is processed as-is. For the binarised RP, the first option with percentage of $1-p$ of pixels is applied.

\begin{figure}
    \centering
    \includegraphics[width=\textwidth]{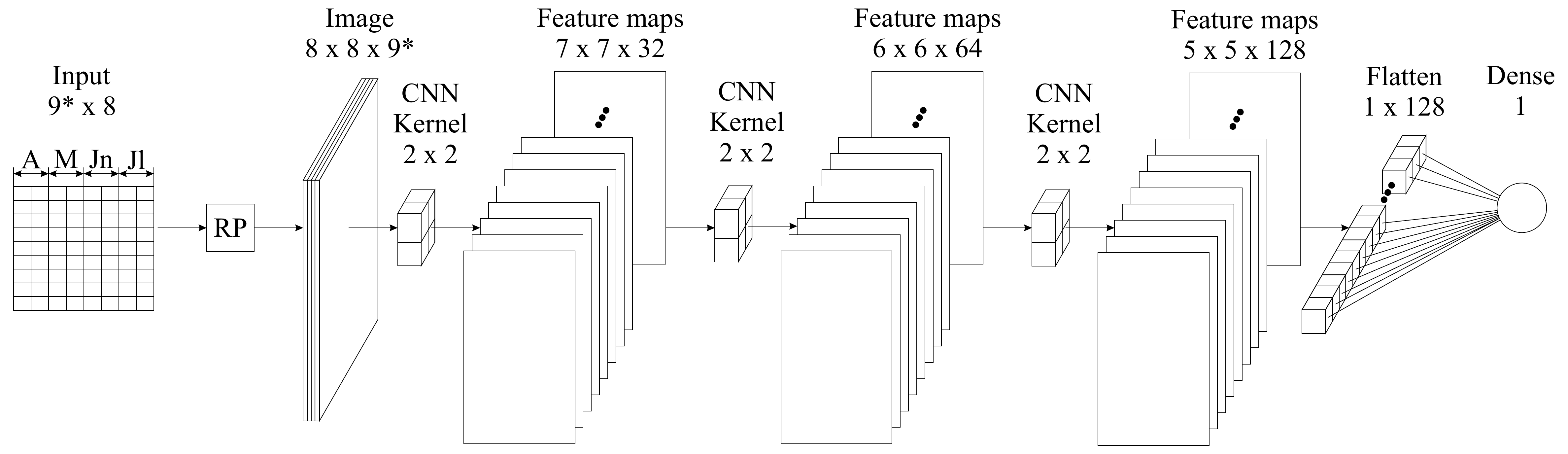}
    \caption{CNN architecture. The input consists of a RPs of dimensions $8 \times 8 \times 9^*$. Months from April to July are depicted on the RPs. The CNN processes the input data using 3 separate 2D kernels, hence 3 sets of feature maps are generated. The output is organised as a combination of a flattened and a dense layer and represents the $y_{t}$ in either the $\tau_1$ or $\tau_2$ prediction horizons. No differences are made to architectures between RP+CNN or RP+CNN(+BIN). Notes: A=April, M=May, Jn=June, Jl=July.
    }
    \label{fig:rp_cnn_arch}
\end{figure}

The RP+CNN(+BIN) architecture is shown in Figure \ref{fig:rp_cnn_arch}. Each input instance is organised tabularly, with dimensions $9^* \times 8$. These are transformed by RP with dimensions $8 \times 8 \times 9^*$. For the CNN, a reduced $2 \times 2$ kernel is employed to process the input images. There are 32 feature maps in the first hidden layer, imitating the CNN computational framework setting. Generated first layer of feature maps is reduced from the $8 \times 8$ to the dimension of $7 \times 7$. Another hidden layer of 64 feature maps follows, again imitating the CNN framework. Final CNN layer is the third layer of 128 feature maps with the dimensions of $5 \times 5$. Feature maps are collected by a single flattening layer with 128 neurons that outputs a single regression value.

\section{Experiments and Results}\label{sec:Experiments}

This experimental section is divided into two subsections, each of them dealing with long-term air temperature forecasts in Paris (France) and Córdoba (Spain), respectively. Two different experiments are conducted for each area, the first one for the shorter ($\tau_1$) prediction time-horizon, and the second one for the prolonged ($\tau_2$) prediction time-horizon. The objective is to forecast the air temperature $\hat{y}_{t}$ in the considered study area (cities) for the given prediction time-horizon with the minimum possible errors (deviations).

The methodology carried out is the following: First, the climate data are obtained, treated and fused, and further adjusted to comply with the specifics of each method (Table \ref{tab:data} shows the input and output data for each of the employed family of methods). Period from April--July is adopted to represent the sequence of input variables, and August as the target month (forecast). In total, 72 years from 1950--2021 are considered in the study, of which 52 instances during 1950--2001 are considered for training, and the rest 20 instances during 2002-2021 as out-of-sample forecasts (test). For each given study area and for each prediction time-horizon, multiple algorithms are tested, in total 9. The first 3 of them belong to the family of deterministic shallow ML methods, the next 3 to the family of the stochastic shallow ML methods, the last 3 are the stochastic CNN methods (for stochastic methods $N=10$ independent runs are considered instead of a single one, to avoid the stochastic bias). In total, $9^*$ predictor variables and a single $y_{t}$ output are supplied to each model. For each method specifically, the EFS procedure is run.

Results are interpreted by a combination of performance graphics and a set of performance metrics. Performance graphics indicate in detail (1) how consistently each method forecasts $\hat{y}_{t}$ with minimum error from actual $y_{t}$; (2) how well each methods adjusts to the trend of slight $y_{t}$ increase within the forecasting period, and (3) how well each method forecasts the $y_{t}$ outliers, i.e., observations far away from long-term average, therefore possibly indicating a heatwave or a coolwave signal appearing in August summer air temperature. Performance metrics are given in numerical values and indicate how well the forecasts are as a whole. For each method, the following metrics are considered: (1) the numeric rank according to the mean squared error statistical indicator, (2) two most common statistical indicators, i.e., $mse$ and mean absolute error ($mae$), (3) two correlation coefficients, i.e., the Pearson and Spearman, with appropriate statistical significance, and (4) the optimal subset of predictor variables obtained by the exhaustive search.

\begin{table}[ht]
    \centering
    \begin{tabular}{lcc} \toprule
    Method     & Input    & Output  \\ \midrule
    ML methods & $x_{tk}''$ &  $y_{t}''$ \\
    CNN     & $x_{ijt}^{(k)''}$ &  $y_{ijt}'''$ \\
    RP+CNN(+BIN)  & $x_{tk}''$ &  $y_{t}''$ \\ \bottomrule
    \end{tabular}
    \caption{Input and output data as required by each of the family of methods.}
    \label{tab:data}
\end{table}

The evaluation function for EFS is defined as $z_{m}=mse$. Evaluation function is adjusted for stochastic models, as follows in Equation \eqref{eq:fitness}, effectively averaging the $mse_{h}$ performance among the $N=10$ runs. Here, the $mse_{h}$ denotes the mean squared error or $mse$ of the $h$-th model, the lower the error, the better the model. Only the best model, according to the best evaluation function value for each method is shown in the results. Parameter settings as outlined in table \ref{tab:exp_setup} were used for modelling methods. Finally, the CNN and RP+CNN(+BIN) architecture settings are listed in Tables \ref{tab:cnn_arch} and \ref{tab:rp_cnn_bin_arch}.

\begin{equation}
    z_{m}=\frac{\sum_{run=1}^{N=10} mse_{h}}{N}.
    \label{eq:fitness}
\end{equation}

\begin{table}[ht]
    \centering
    \footnotesize
    \begin{tabular}{lr} \toprule
    Variable     &  Setting \\ \midrule
    Paris geographical coordinates* & 48.75$^\circ$N, 2.25$^\circ$E \\
    Córdoba geographical coordinates* & 37.75$^\circ$N, 4.75$^\circ$W \\ \midrule
    LR's learning algorithm & OLS \\
    Lasso's $\lambda$ param & 0.0005 \\
    No. of polynomial degrees & 4 \\
    AdaBoost's no. of estimators & 100 \\
    DT's max. depth & 10 \\
    RF's max. depth & 10 \\ \midrule
    Learning algorithm of the CNN, RP+CNN(+BIN)    & Adam \cite{kingma2014adam} \\
    Learning rate of the CNN, RP+CNN(+BIN) & 0.001 \\
    \hline
    \end{tabular}
    \caption{Parameter settings for the modelling methods. *=rounded to nearest quarter. The first box exhibits basic information; the second box shows the ML experimental setup. Third box exposes CNN setup and the fourth shows the RP+CNN(+BIN) settings.}
    \label{tab:exp_setup}
\end{table}

\begin{table}[ht]
    \centering
    \begin{tabular}{lrrr} \toprule
    Block type  & Ingredients  & Kernel size & Size of feature maps \\ \midrule
    input & ~ & ~ & 8 $\times$ $33 \times 33 \times 9^*$ \\
    down 1 & Conv3D/relu & $3 \times 3 \times 3$ & $6 \times 31 \times 31 \times 32$ \\
    down 2 & Conv3D/relu & $3 \times 3 \times 3$ & $4 \times 29 \times 29 \times 64$ \\
    down 3 (output) & Conv3D/sigmoid & $4 \times 3 \times 3$ & $1 \times 27 \times 27 \times 1$ \\ \bottomrule
    \end{tabular}
    \caption{CNN architecture. Number of channels ($9^*$) are subject to change due to the exhaustive search.}
    \label{tab:cnn_arch}
\end{table}

Kernel size settings were set to minimal values practicable, as suggested by \cite{simonyan2014very}, who realised that a very small kernel size, e.g., $3 \times 3$, delivers significant improvements and increases the CNN effectiveness. Additionally, small kernel size has also been used because of the low input image size dimension, which has been selected due to the geographical constraints, namely the homogeneous area with relatively uniform correlation coefficients. Furthermore, the kernel size has been further reduced to $2 \times 2$ in case of RP+CNN(+BIN) due to very small input image size dimension ($8 \times 8$). Due to operating with very small input image size dimensions on one hand, but higher number of channels on the other, no pooling layers to reduce the dimensionality have been introduced to any framework. The introduction of the Experiments and results section is finalised by the Algorithm \ref{pseudocode} representing the pseudocode of the $\hat{y}_{t}$ forecasts.

\begin{table}[ht]
    \centering
    \begin{tabular}{lrrr} \toprule
    Block type  & Ingredients  & Kernel size & Size of feature maps \\ \midrule
    input & ~ & ~ & $8 \times 8 \times 9^*$ \\
    down 1 & Conv2D/relu & $2 \times 2$ & $7 \times 7 \times 32$ \\
    down 2 & Conv2D/relu & $2 \times 2$ & $6 \times 6 \times 64$ \\
    down 3 & Conv2D/relu & $2 \times 2$ & $5 \times 5 \times 128$ \\
    output & Flatten \& Dense/linear & 1 & 1 \\ \bottomrule
    \end{tabular}
    \caption{RP+CNN(+BIN) architecture. Number of channels ($9^*$) are subject to change due to the FS.}
    \label{tab:rp_cnn_bin_arch}
\end{table}

\begin{algorithm}[ht]
\begin{algorithmic}[1]
\Procedure{Forecasting the $\hat{y}_{t}$ using the ML, CNN, RP+CNN(+BIN)}{}
 \State INITIALISE city and time horizon;
 \State $x_{t}^{(k)''}, x_{ijt}^{(k)''} \gets$ FUSE and ADJUST the input data;
 \State $y_{t}'', y_{ijt}'' \gets$  FUSE and ADJUST the output data;
 \State $u \gets$ GENERATE all possible combinations of predictor variables;
 \For{all possible combinations $\textbf{u}$}
 \For{all modelling methods $\textbf{g}$}
  \State TRAIN MODEL on subset of predictors $u_n$ for model $g_m$;
  \State MAKE FORECASTS $\hat{y}_t$ on the trained model $g_m$;
  \State $z_{n,m} \gets$ CALCULATE $mse$ for the subset $u_n$ for model $g_m$;
  \EndFor
 \EndFor
\EndProcedure
\end{algorithmic}
\caption{The pseudocode of the temperature forecasts in a given study area.}
\label{pseudocode}
\end{algorithm}

This pseudocode shows the workflow of the $\hat{y}_{t}$ forecasts for a given study area and a given prediction time-horizon. After the study area and the prediction time-horizon are defined, the input and output data are fused and adjusted to comply with the requirements of each specific method. Then, the EFS is run for each modelling method. Each possible combination of predictors is sequentially trained and forecast is obtained. The deterministic models run the trial solutions just a single time, others $N=10$ times. For the latter, an average of $mse_{h}$ is calculated to evaluate the quality of forecasts. Finally, a vector of new trial solutions $\textbf{u}_{m}$ is generated. The iterative procedure is run until the stopping criteria is met, i.e., the number of function evaluations hit the $nFEs\_max$. The next subsection reports the results on forecasting the $\hat{y}_{t}$ in Paris.

Code was written exclusively in Python programming language. Data fusioning, adjusting and handling were done with the following Python libraries: Pandas \cite{reback2020pandas,mckinney-proc-scipy-2010}, Numpy \cite{harris2020array} and Xarray \cite{xarray}. RPs were created in Pyts \cite{JMLR:v21:19-763}. For the implementation of ML methods, sklearn \cite{scikit-learn} library was chosen. The CNN architectures were implemented with Keras \cite{chollet2015keras} and Tensorflow \cite{tensorflow2015-whitepaper} libraries.

\subsection{Results for long-term temperature forecasts in Paris}

This subsection starts with the comment on the $y_{t}$ dynamics for Paris in years 2002--2021, and continues with the performance graphics. Results on nine different modelling methods are visualised in a shape of a $3 \times 3$ table. Later, the performance metrics with a set of five statistical indicators and a best EFS combination follow.

\begin{figure}
\centering
\begin{subfigure}{1.0\textwidth}
    \includegraphics[width=\textwidth]{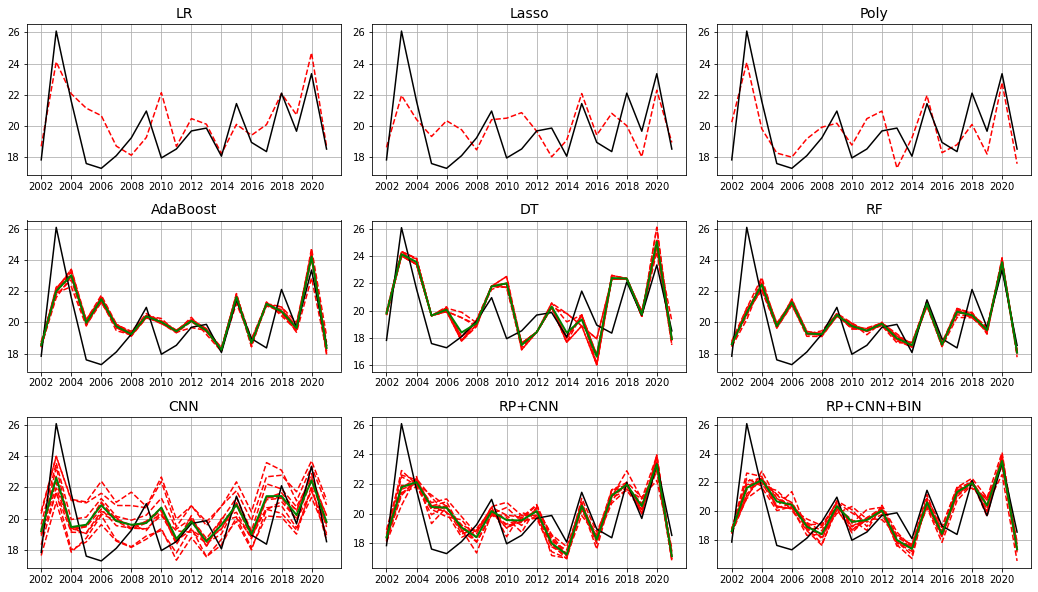}
    \caption{Temperature forecasts $\hat{y}_{t}$ in Paris, $\tau_1$.}
    \label{fig:paris_short}
\end{subfigure}
\hfill
\begin{subfigure}{1.0\textwidth}
    \includegraphics[width=\textwidth]{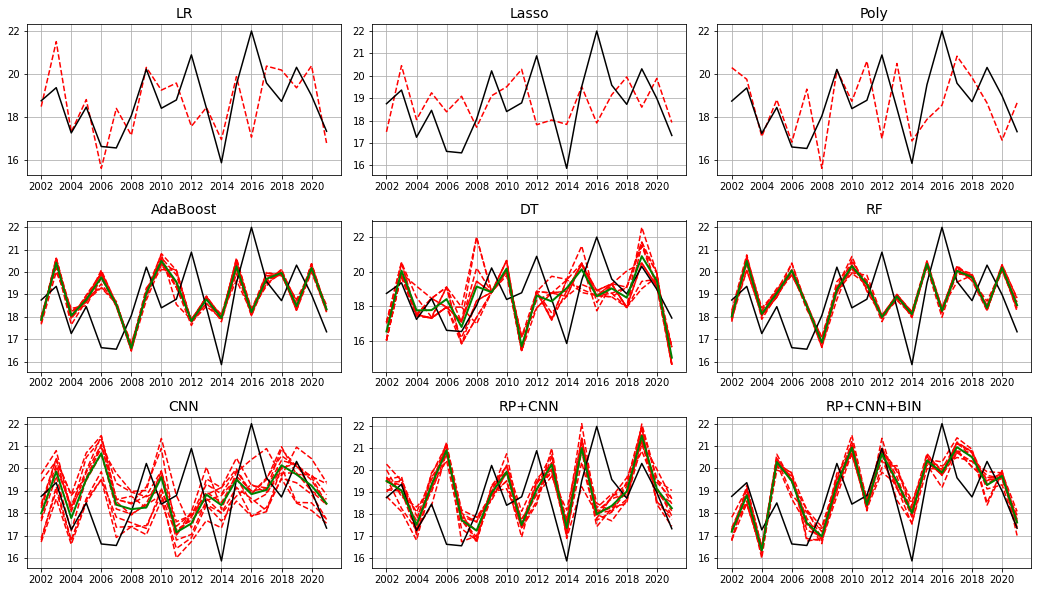}
    \caption{Temperature forecasts $\hat{y}_{t}$ in Paris, $\tau_2$.}
    \label{fig:paris_long}
\end{subfigure}
\caption{Forecast of average daily mean temperature in August ($\hat{y}_{t}$) in Paris. Solid black line represents the true $y_{t}$ in Paris, red dotted lines represent individual runs of $\hat{y}_{t}$, solid green represents the average of the individual runs (not applicable for deterministic models in first row). The first row represents the deterministic ML methods, LR, Lasso and Polynomial regressions. Second row shows results for more complex ML methods, such as AdaBoost, DT and RF. The third row shows the results of the proposed methodologies, CNN, RP+CNN and the RP+CNN+BIN.}
\label{fig:paris}
\end{figure}

The daily mean air temperature ($y_{t}$ ) in Paris in August ranges from $17.43^\circ$ to $26.61^\circ$ Celsius, with a mean of $19.94^\circ$C and a variance $4.96^\circ$C. Period within 2002--2009 shows a very unsteady and difficult-to-predict $y_{t}$ performance, associated with an extreme event in year 2003. The temperature rises rapidly during one year, which is then followed by approximately 3 years of $y_{t}$ lower than usual. Since 2009, the time series is more stable, quasi first-order negatively autocorrelated. Therefore, we expect a worse performance in the first part of the time series and a better performance in the second. Figure \ref{fig:paris_short} shows the performance graphics of forecasting $\hat{y}_{t}$ on shorter prediction time-horizon (first fortnight of August), and Figure \ref{fig:paris_long} on prolonged prediction time-horizon (second fortnight of August).

Interpretation of the modelling methods is as follows. For shorter time horizon, all the methods included exhibit underestimations during the extreme weather event in year 2003 for shorter-time horizon $\tau_1$. All of them also underestimate the temperature drop during 2006 cool event. Contrary, all methods indicate the temperature increases in 2020 well. Visually, Poly is the best fit among modelling methods in the horizon $\tau_1$, since it best forecasts the 2003 year heatwave and associated temperature drop afterwards. It delivers the best compromise between forecasts during non-extreme (regular, typical, casual) events and forecasts during extreme events. ML methods show a lower level of variability than CNN-based methods. Among them, RP+CNN+BIN is the most promising by visual means, since it delivers the best compromise between variability and non-extreme events forecasting.

Visually, for the prolonged horizon $\tau_2$, RP+CNN and RP+CNN+BIN seem to be the best fit. Predictions are less variable than for the horizon $\tau_1$. This is positive, but lower variability inherently implies lower skill on forecasting extremes. Deterministic and ML methods lack of forecast skills in years 2005 and 2016. ML techniques also lack of forecast skill in years 2011 and 2014. CNN-based methods are far from perfect, but capture the trend and magnitudes to the best degree among all methods analysed. We deduce that the more complex the modelling method, the better the forecast for prolonged time horizon.

\begin{table}[!ht]
    \centering
    \begin{tabular}{rrrrrrr} \toprule
        \multicolumn{7}{c}{Paris $\tau_1$}  \\ \toprule
        ~ & rank & $mse$ & $mae$ & Pearson & Spearman & Vars. \\
        LR         & 3(9) & 2.973 & 1.264 & **0.695 & 0.409  & 011000100  \\
        Lasso      & 7(6) & 3.316 & 1.519 & *0.559  & 0.347  & 111100000  \\
        Poly       & 1(7) & 1.928 & 1.226 & **0.772 & *0.535 & 101100100  \\
        AdaBoost   & 5(1) & 3.092 & 1.236 & **0.645 & *0.519 & 111111100  \\
        DT         & 9(4) & 3.651 & 1.508 & **0.665 & 0.433  & 010100010  \\
        RF         & 8(2) & 3.375 & 1.247 & *0.559  & 0.424  & 110100100  \\
        CNN        & 6(5) & 3.232 & 1.446 & **0.606 & 0.292  & 100000001  \\
        RP+CNN     & 2(8) & 2.970 & 1.332 & **0.63  & *0.517 & 110000100  \\
        RP+CNN+BIN & 4(3) & 3.007 & 1.295 & **0.625 & *0.462 & 000001010  \\ \toprule
        \multicolumn{7}{c}{Paris $\tau_2$}  \\ \toprule
        ~ & rank & $mse$ & $mae$ & Pearson & Spearman & Vars. \\
        LR         & 2(9) & 2.704 & 1.159 & 0.401  & *0.526 & 100010010  \\
        Lasso      & 3(6) & 2.722 & 1.333 & 0.121  & 0.177  & 100100010  \\
        Poly       & 9(8) & 3.299 & 1.475 & 0.253  & 0.25   & 101100010  \\
        AdaBoost   & 6(2) & 3.062 & 1.457 & 0.091  & 0.116  & 100001011  \\
        DT         & 5(7) & 2.938 & 1.362 & 0.343  & 0.365  & 010101100  \\
        RF         & 7(3) & 3.064 & 1.478 & 0.081  & 0.123  & 100001011  \\
        CNN        & 8(4) & 3.072 & 1.348 & -0.01  & 0.041  & 111110000  \\
        RP+CNN     & 4(5) & 2.931 & 1.351 & 0.246  & 0.286  & 010010101  \\
        RP+CNN+BIN & 1(1) & 2.117 & 1.261 & *0.516 & *0.487 & 100000001  \\ \bottomrule
    \end{tabular}
    \caption{Statistical indicators of Paris $\tau_1$ and $\tau_2$ forecasts. Ranks in brackets represent the non-FS ranks (all predictor variables included). "Pearson/Spearman"=Pearson's and Spearman's rank correlation coefficients, *=$p$-value less than 0.05, **=$p$-value less than 0.01, "Vars."=variables ordered as \{t2m, u10, u100, v10, v100, msl, sst, geo500, swvl1\}. Ranks calculated on the basis of $mse$ value.}
    \label{tab:paris}
\end{table}

Table \ref{tab:paris} represents performance metrics of forecasts in Paris, for both $\tau_1$ and $\tau_2$. Poly is the best modelling method according to the performance metrics for shorter prediction horizon and RP+CNN+BIN for prolonged horizon $\tau_2$. Both of them are significantly better than the rest of the methods, regarding the $mse$ statistical indicator. Correlation coefficients are significant for all methods for shorter prediction and significant only for LR and RP+CNN+BIN for prolonged horizon (for either Pearson's or Spearman's coefficients). The use of EFS drastically lowers the number of predictors, e.g. RP+CNN+BIN only includes two variables. As expected, the air temperature predictor seems to be among the more important.

\subsection{Results for long-term temperature forecasts in Córdoba}

Average daily mean August air temperature in Córdoba ranges from $25.78^\circ$ to $29.87^\circ$ Celsius, with a mean $27.66^\circ$C and a variance $1.02^\circ$C. Two extreme temperature events are spotted in the test period considered, one in the famous 2003 summer, the other in years 2017--2018. A significant cool event is spotted in year 2014. Córdoba experimented a gradual increase in temperatures in years 2002--2021, which even more intensifies the challenge of forecasting.

\begin{figure}
\centering
\begin{subfigure}{1.0\textwidth}
    \includegraphics[width=\textwidth]{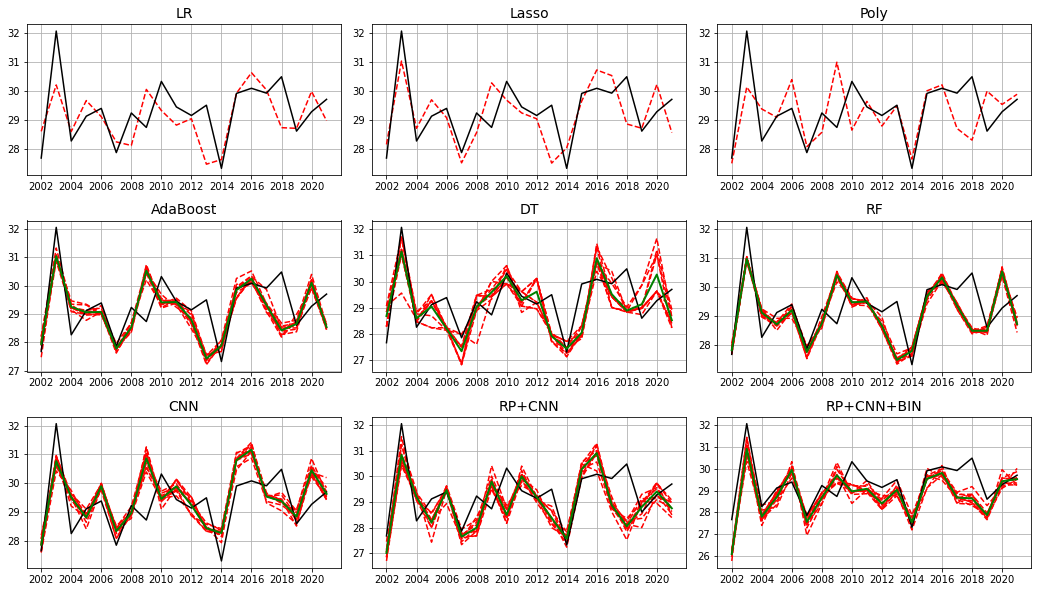}
    \caption{Temperature forecasts $\hat{y}_{t}$ in Córdoba, $\tau_1$.}
\end{subfigure}
\hfill
\begin{subfigure}{1.0\textwidth}
    \includegraphics[width=\textwidth]{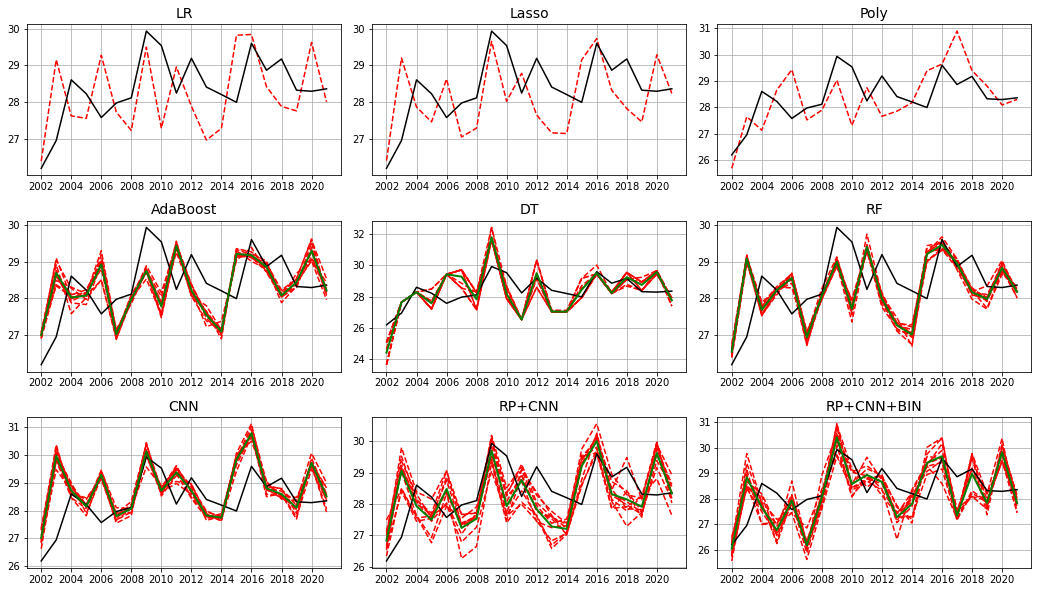}
    \caption{Temperature forecasts $\hat{y}_{t}$ in Córdoba, $\tau_2$.}
\end{subfigure}
\caption{Forecasting the $y_{t}$ in Córdoba. Solid black line represents the true $y_{t}$ in Córdoba, red dotted lines represent individual runs of $\hat{y}_{t}$, solid green represents the average of the individual runs (not applicable for models in first row). The first row represents the deterministic ML methods, LR, Lasso and Polynomial regressions. Second row represents results of more complex ML methods, such as AdaBoost, DT and RF. The third row represents the results of the proposed methodologies, CNN, RP+CNN and the RP+CNN+BIN. Years on the x-axis, 2 meter temperature in $^\circ$ Celsius on the y-axis.}
\label{fig:cordoba}
\end{figure}

Performance graphics are visualised in Figure \ref{fig:cordoba}. First impression is that the Córdoba is more forecastable than Paris area. By far, the best forecasts are in this case provided by the RP+CNN+BIN. With the exception of years 2010--2013 and years 2017--2018, forecasts are very similar to the actual temperatures, either at extreme or non-extreme events. Among the deterministic methods, Lasso is the best compromise. CNN variability is much decreased compared to the Paris case, which is again a sign that different study areas have different forecastibilities. Despite, all the methods are prone to the erroneous forecasting in years 2017--2018.

Performance metrics can be found in Table \ref{tab:cordoba}. RP+CNN+BIN is found to be the best method for shorter, RP+CNN for prolonged forecast horizon. Compared to the Paris, $mse$ of both horizons are decreased much and correlation coefficients are increased. Three of the Pearson's coefficients are significant for methods during the prolonged forecast horizon. It is realised that Poly does not deliver stable performance, since ranks are inverted compared to the Paris and correlation coefficients are insignificant. EFS again reduces much the sets of most suitable predictors.

\begin{table}[!ht]
    \centering
    \begin{tabular}{rrrrrrr} \toprule
        \multicolumn{7}{c}{Córdoba $\tau_1$}  \\ \toprule
        ~ & rank & $mse$ & $mae$ & Pearson & Spearman & Vars. \\
        LR &        6(9) & 0.903 & 0.740 & *0.548  & *0.531 & 100000000 \\
        Lasso &     2(7) & 0.778 & 0.720 & **0.639 & *0.538 & 100000100 \\
        Poly &      9(8) & 1.149 & 0.770 & 0.423 & 0.332    & 100111010 \\
        AdaBoost &  7(3) & 0.911 & 0.705 & **0.577 & 0.441  & 101010100 \\
        DT &        4(2) & 0.833 & 0.745 & **0.636 & 0.432  & 111010100 \\
        RF &        5(1) & 0.872 & 0.712 & **0.604 & *0.483 & 101010100 \\
        CNN &       3(4) & 0.814 & 0.741 & **0.614 &**0.568 & 111000010 \\
        RP+CNN &    8(6) & 1.029 & 0.808 & **0.609 & *0.486 & 000100111 \\
        RP+CNN+BIN & 1(5) & 0.718 & 0.696 & **0.789& **0.651& 000110100 \\ \toprule
        \multicolumn{7}{c}{Córdoba $\tau_2$}  \\ \toprule
        ~ & rank & $mse$ & $mae$ & Pearson & Spearman & Vars. \\
        LR &        9(9) & 1.398 & 1.001 & 0.230 & 0.180    & 000100000 \\
        Lasso &     5(7) & 1.093 & 0.906 & 0.364 & 0.331    & 000001110 \\
        Poly &      6(8) & 1.094 & 0.791 & *0.454 & 0.257   & 101001010 \\
        AdaBoost &  2(2) & 0.976 & 0.841 & 0.249 & 0.165    & 111100000 \\
        DT &        8(6) & 1.196 & 0.914 & **0.678 & 0.441  & 000000010 \\
        RF &        3(1) & 1.012 & 0.822 & 0.313 & 0.164    & 100101000 \\
        CNN &       7(4) & 1.145 & 0.755 & 0.396 & 0.322    & 010100100 \\
        RP+CNN &    1(5) & 0.971 & 0.857 & 0.373 & 0.314    & 010101000 \\
        RP+CNN+BIN & 4(3) & 1.028 & 0.829 & *0.538 & 0.380  & 111011111 \\ \bottomrule
    \end{tabular}
    \caption{Statistical indicators of Córdoba $\tau_1$ and $\tau_2$ forecasts. Ranks in brackets represent the non-FS ranks (all predictor variables included). "Pearson/Spearman"=Pearson's and Spearman's rank correlation coefficients, *=$p$-value less than 0.05, **=$p$-value less than 0.01, "Vars."=variables ordered as \{t2m, u10, u100, v10, v100, msl, sst, geo500, swvl1\}. Ranks calculated on the basis of $mse$ value.}
    \label{tab:cordoba}
\end{table}

\section{Conclusions}\label{sec:Conclusions}
Seasonal climate prediction problems involve uncertain and demanding tasks related to forecasting the long-term steady-levels of different climate variables, such as air temperature. In this long-term behaviour of variables, it is possible to spot short-term extreme events signals, such as heatwaves. Some geographical areas are in fact more exposed to weather extremes than others, and hence, these extreme signals should appear in the long-term prediction of climate variables at these zones. In line with this, no universal model can fit forecasts for all the geographical areas well, which means that not only the span of coordinates of input data may be different to forecast in a specific area, but also the set of the best input data features (data variables) may be different. 

Following this idea, in this paper we have tackled a problem of long-term air temperature prediction in summer (August), using different computational frameworks based on AI techniques. Specifically we first propose a novel approach based on CNN combined with different process for data fusion and dimensionality reduction. In the second computational framework, different ML approaches are proposed, including Lasso, regression trees and Random Forest. The third computational framework also considers a CNN, with pre-processing steps via RPs as data reduction technique. RPs have been assimilated as a compromise to exploit the temporal structure of the data. Since the RP is a transformation of a time-series into an image, the CNN has been further exploited in this case with the RPs output as a processing medium. 

The performance of the different proposed AI-based computational frameworks have been evaluated in two problems of long-term air temperature prediction at Paris (France) and Cordoba (Spain), considering the prediction in the first and second August fortnights using predictive variables from the previous months. The results obtained seem to indicate a superior performance by the RP+CNN-based approaches, albeit no unique model is the best approach for both prediction time-horizons considered. The proposed RP+CNN-based approaches were able to accurately detect some maximums in the summer temperature better than classical CNN and ML techniques. These maximum values can be associated with heatwaves signals occurring in August in the areas studied (Paris and Córdoba), such as that of 2003, whose signal is detectable in the August mean temperature when comparing with other years. 

As future research lines, we propose that the original CNN model could be reworked to output not only a single channel (like the 2 meter air temperature in this paper), but rather a set of multiple channels, including the wind information and/or volumetric soil water layers. Increased complexity due to multi outputs could be compensated by data augmentation techniques to achieve identical stability of the models. Different architectures, including auto-encoders, could be employed to exploit the benefit of converging the images into a single-size and diverging it back to the original size. In general, a larger amount of climate data could be exploited for training the models, by including the climate data from January--April and from September--December, to better capture the climate trend. Also, a universal model that would fit forecasts for all geographical areas should be built and verified compared to the ML and DL methods.

\section*{Open Research}
All the data used in this paper are from ERA5 Reanalysis, available under request to the European Centre for Medium-Range Weather Forecasts  \cite{ECMWF}.

\section*{Acknowledgement}
This research has been partially supported by the European Union, through H2020 Project ``CLIMATE INTELLIGENCE Extreme events detection, attribution and adaptation design using machine learning (CLINT)'', Ref: 101003876-CLINT. This research has also been partially supported by the project PID2020-115454GB-C21 of the Spanish Ministry of Science and Innovation (MICINN). Javier Del Ser is supported by the Basque Government through the ELKARTEK program and the consolidated research group MATHMODE (IT1456-22)

\bibliography{agujournaltemplate} 
%




%
%
%
%
%

\end{document}